\documentclass[12pt]{iopart}
\usepackage[normalem]{ulem}
\usepackage{multirow}
\usepackage[utf8]{inputenc}
\usepackage{iopams}
\usepackage{hyperref}
\usepackage{color}  
\expandafter\let\csname equation*\endcsname\relax
\expandafter\let\csname endequation*\endcsname\relax
\usepackage{mathtools}
\usepackage[]{graphicx}
\usepackage{multicol}
\graphicspath{{./}
	{./figsPng/}
}
\begin{document}
	%***************************************************************************************
	
\title{
	Instabilities and turbulence in stellarators from the perspective of global  codes
}
\author{E. Sánchez$^1$, A. Bañón Navarro$^2$, F. Wilms$^2$, M. Borchardt$^3$, R. Kleiber$^3$, F. Jenko$^2$}
%  \corresp{\email{edi.sanchez@ciemat.es}},
%
\address{$^1$Laboratorio Nacional de Fusi\'on - CIEMAT. Avda. Complutense 40, 28040, Madrid, Spain.}
\address{$^2$Max-Planck Insitut f\"ur Plasmaphysik, Garching, Germany.}
\address{$^3$Max-Planck Insitut f\"ur Plasmaphysik, Greifswald, Germany.}
\ead{edi.sanchez@ciemat.es}
\vspace{10pt}

%\maketitle

%***************************************************************************************
\begin{abstract}
	In this work, a comparison of the global gyrokinetic codes EUTERPE and GENE-3D in stellarator configurations of LHD and W7-X is carried out. In linear simulations with adiabatic electrons, excellent agreement is found in the mode numbers, growth rate and frequency, mode structure, and spatial localization of the most unstable mode in LHD. In W7-X, %the agreement between codes in single simulations is not as good as in LHD but 
	the dependence of the growth rate and frequency with the mode number is well reproduced by both codes. The codes are also compared in linear simulations with kinetic ions and electrons in W7-X using model profiles, and reasonable agreement is found in the wavenumber of the most unstable modes. A stabilization of small-scale modes in kinetic-electron simulations with respect to the adiabatic-electron case is consistently found in  both codes.
	 Nonlinear simulations using adiabatic electrons and model profiles are also studied and the %\pensar{spectrum and} 
	 heat fluxes are compared. Very good agreement is found in the turbulent ion heat fluxes in both LHD and W7-X. Two problems that cannot be properly accounted for in local flux tube codes are studied: the localization of instabilities and turbulence over the flux surface and the influence of a background long-wavelength electric field. Good agreement between codes is found with respect to the spatial localization of instabilities and turbulence over the flux surface. The localization of saturated turbulence is found in both codes to be much smaller than that of the linear instabilities and smaller than previously reported in full-surface radially-local simulations. The influence of the electric field on the localization is also found to be smaller in the developed turbulent state that in the linear phase, and smaller than in previous works. A stabilizing effect of a constant electric field on the linearly unstable modes is found in both codes. A moderate reduction of turbulent transport by the radial electric field, with small dependence on the sign of the electric field, is also found.
\end{abstract}
%
%
%***************************************************************************************
\section{Introduction}
%***************************************************************************************
Many codes have been developed based on the gyrokinetic formalism \cite{CATTO78,FRIEMAN82,BRIZARD07, PARRA11} and have been used for the simulation of plasma turbulence in toroidal devices. 
Simulation codes can result extremely useful for the understanding of plasma turbulence, as it is a complex problem with no possible fully analytic treatment, except in simplified or particular cases.  
The physical problem can be treated numerically by using different numerical implementations, each of them having their own weakness and strengths. Besides, these kinds of numerical models imply some approximations and simplifications, 
which makes it very important to verify the numerical codes by means of comparisons to analytical models when possible, or against other codes with different numerical implementations.

The axial symmetry in tokamaks makes it possible to use of the local, so-called flux tube approximation \cite{Beer95},  consisting of the simulation of a physical domain surrounding a magnetic field line and following the line one poloidal turn, which allows reducing significantly the computational resources required for turbulence simulations with these codes with respect to the simulation of either the full domain or the full flux surface. In stellarators, the situation is quite different and different flux tubes lying over the same flux surface cannot be considered as equivalent. Although first applications of gyrokinetic codes to stellarators were based on the direct adaptation of the flux tube paradigm for stellarators \cite{Watanabe2002,Xanthopoulos07,Baumgaertel2011} and the use of flux tubes with just one poloidal turn in length, this is not satisfactory in stellarators. The minimum computational domain required for stellarators was addressed in recent works studying two linear problems \cite{Smoniewski19,SanchezNF2021} and it was demonstrated that, at fixed radial wavenumber, short flux tubes on the same flux surface provide different results, in general, which only converge to each other when the flux tube length is sufficiently increased. In \cite{Banon20} it was shown that the heat flux computed in full-surface simulations can be significantly larger than that obtained in full global simulations, suggesting that global simulations are required in stellarators, in general. Furthermore, global codes are, in principle, required in stellarator turbulence simulations to properly account for the influence of the long-wavelength electric field and the global density and temperature profiles.

Several global gyrokinetic codes have been specifically designed for stellarators or adapted from tokamak codes for thee dimensional geometries, and there is presently a number of them available for stellarators: 
XGC-S \cite{Cole19}, GTC \cite{Wang20}, GT5D \cite{Matsuoka18}, EUTERPE \cite{Jost01,Kornilov04}, GENE-3D \cite{Maurer19} and GKNET \cite{Imadera14}. 
While in tokamaks gyrokinetic simulations have a reasonable degree of maturity and there is a set of codes cross-benchmarked and validated, the number of available codes able to target the stellarator geometry has been  limited until recently, and they still lack of verification and validation, in general.  
In this work, we present the results of a effort carried out during the last months for the cross-verification of the codes EUTERPE \cite{Jost01,Kornilov04} and GENE-3D \cite{Maurer19}. Both are global codes designed specifically for stellarator geometry. They are based on different numerical models but they also share common features that permit an in-detail comparison from which both codes can benefit. In addition to the cross verification of these codes, we address two problems that are considered relevant for turbulent transport in stellarators and for which the global codes are particularly suited, namely the localization of instabilities and turbulence over the flux surface and the influence of the electric field on the instabilities and turbulence.

The structure of the work is as follows. In Section \ref{secCodesyConfs}, we briefly describe the codes under comparison, their numerical models and capabilities and also the stellarator magnetic configurations that we use in this work. In Section \ref{secCompLinearSims}, we compare both codes in linear simulations of Ion-Temperature-Gradient-driven modes (ITGs). The codes are compared in a nonlinear setting in Section \ref{secNonLin}. In Section \ref{secLocalization}, we study the localization of instabilities and turbulence over the flux surface with both codes. Section \ref{secElecField} is devoted to study the influence of a radial electric field on the localization of instabilities and turbulence, the stabilization of linear modes and its effect on the turbulent transport.  Finally, in Section \ref{secSumConcl}, we draw some conclusions.

%%%%%%%%%%%%%%%%%%%%%%%%%%%%%%%%%%%%%%%%%%%%%%%%%%%%%%%%%%%%%%%%%%
%%
 				\section{The magnetic configurations and codes used}\label{secCodesyConfs}
%%
%%%%%%%%%%%%%%%%%%%%%%%%%%%%%%%%%%%%%%%%%%%%%%%%%%%%%%%%%%%%%%%%%%

%%%%%%%%%%%%%%%%%%%%%%%%%%%%%%%%%%%%%%%%%%%%%%%%%%%%%%%%%%%%%%%%%%%%
%%
\subsection{Magnetic configurations}\label{secMagConfs}
\begin{figure}[!h]
	\centering
	\includegraphics[ angle=0,trim=500 190 430 170, clip, width=7.5cm ]{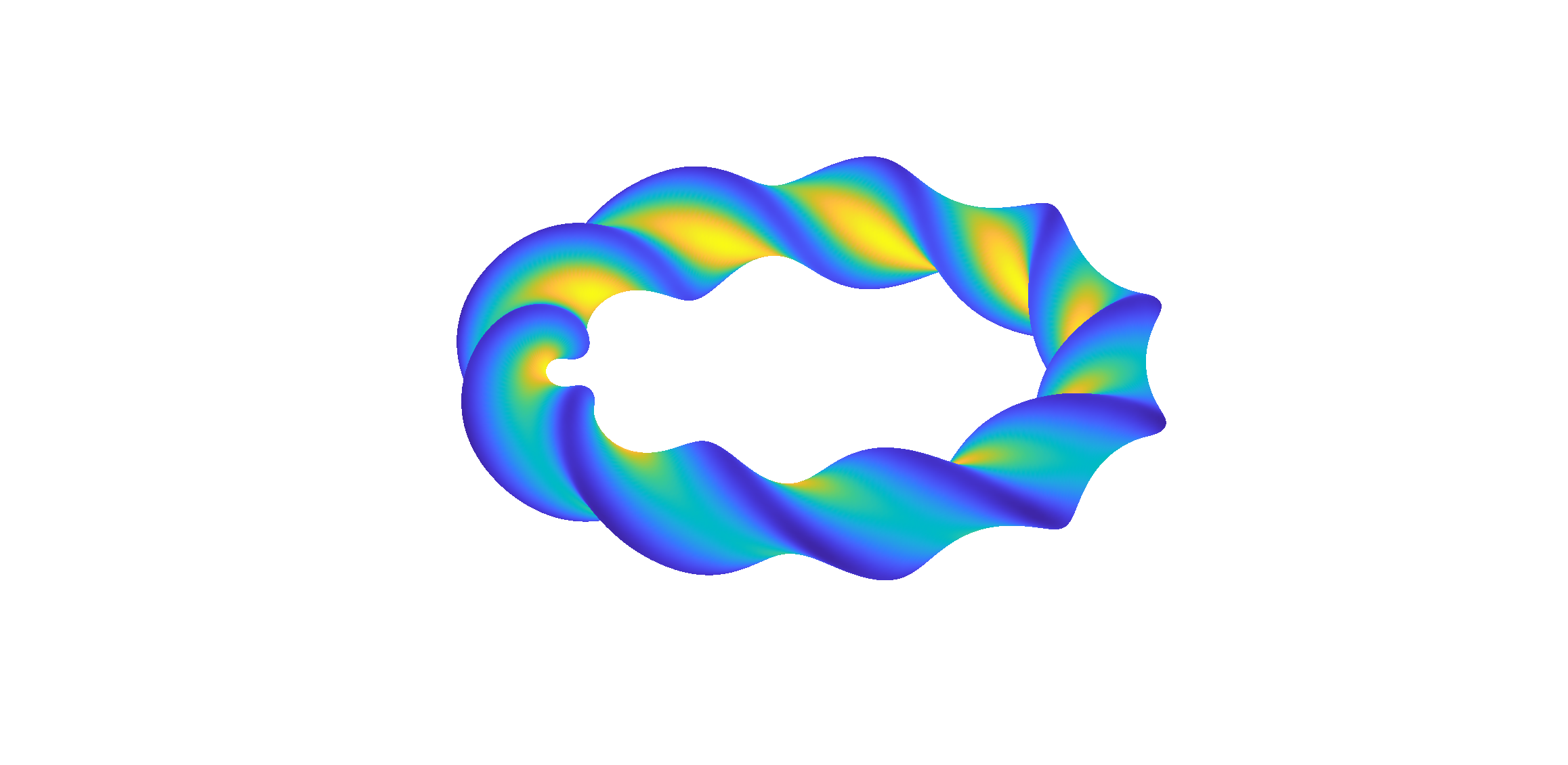}
	\includegraphics[angle=0,trim=420 190 390 175, clip,width=7.5cm]{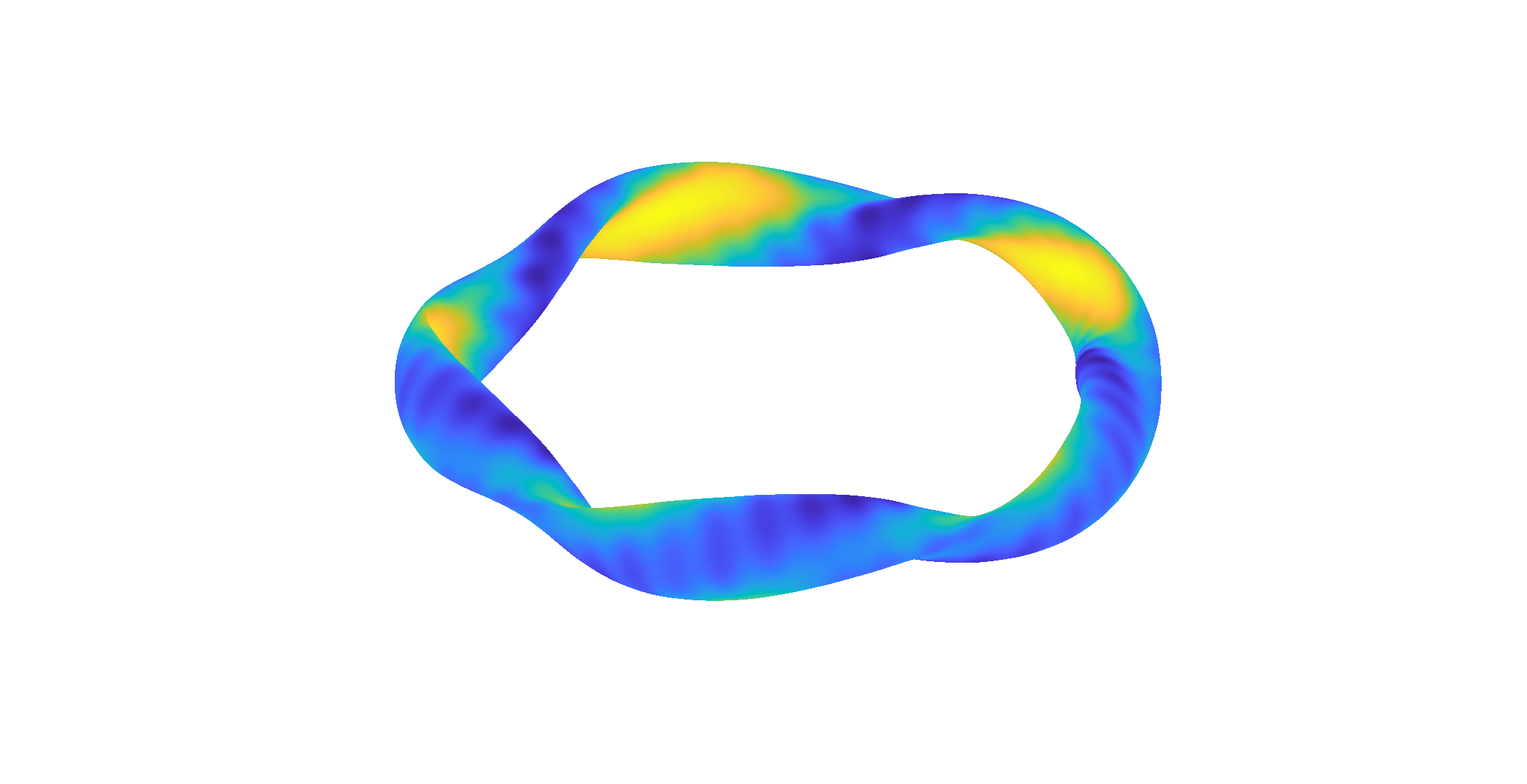}
	\caption{Magnetic field strength on the last closed flux surface for the two magnetic configurations used in this work: a standard configuration of LHD (left) and a standard configuration of W7-X (right).}
	\label{fig:LHDyW7XConfs}
\end{figure}

\begin{figure}[!h]
	\centering
	\includegraphics[width=10cm, angle=0]{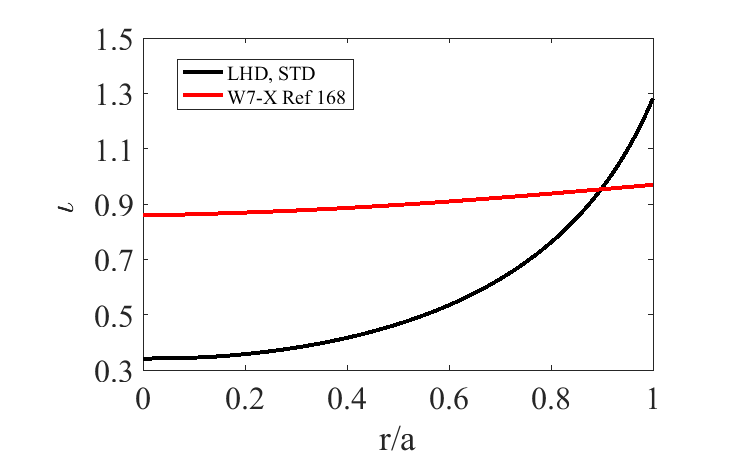}
	\caption{Radial profiles of rotational transform ($\iota$) for the standard magnetic configurations of LHD and W7-X used in this work.}
	\label{fig:iotaLHDyW7X}
\end{figure}

%%
%%%%%%%%%%%%%%%%%%%%%%%%%%%%%%%%%%%%%%%%%%%%%%%%%%%%%%%%%%%%%%%%%%
In this work, we compare simulations carried out in two different stellarator magnetic configurations: a standard configuration of LHD,  with major radius, $R=3.7 ~m$ and minor radius, $a=0.6~m$, and a standard configuration (Ref 168) of W7-X with $R=5.5 ~m$, $a=0.52~m$. The magnetic field strength over the last closed flux surface of both configurations is shown in Figure \ref{fig:LHDyW7XConfs}.  Figure \ref{fig:iotaLHDyW7X} shows the radial profiles of rotational transform in these configurations. 

%*******************************************
%*******************************************

%*******************************************
%*******************************************
LHD has a small rotational transform $\iota=0.3$ at the center that increases toward the edge up to $\iota=1.25$, thus having a significant magnetic shear $\hat{s}=0.08-0.17$,  while for W7-X the rotational transform is in the range $0.86 <\iota < 0.97$ with very low magnetic shear, $\hat{s}=0.0015$, in all  radii (see Figure \ref{fig:iotaLHDyW7X}). 
In both configurations we consider a vacuum equilibrium which is calculated with the code VMEC.% and mapped to internal coordinates of EUTERPE and GENE-3D with VM2MAG and GVEC, respectively.

%%%%%%%%%%%%%%%%%%%%%%%%%%%%%%%%%%%%%%%%%%%%%%%%%%%%%%%%%%%%%%%%%%%%
%%
\subsection{The codes EUTERPE and GENE-3D}\label{secCodes}
%%
%%%%%%%%%%%%%%%%%%%%%%%%%%%%%%%%%%%%%%%%%%%%%%%%%%%%%%%%%%%%%%%%%%
 In this section we briefly describe the codes EUTERPE \cite{Jost01,Kornilov04} and GENE-3D \cite{Maurer19,Wilms21}, their numerical models and their capabilities, and highlight the differences that are relevant for this comparison. Both are global and $\delta$f codes allowing electrostatic or electromagnetic global simulations in stellarators.

  	In this work, the gyrokinetic equation in the collsionless limit,
  	\begin{equation}
  	\frac{\partial F_{\sigma}}{\partial t} + \frac{\partial {\vec{R}}}{\partial t}  \nabla F_{\sigma} + \frac{\partial v_{\|}}{\partial t} \frac{\partial F_{\sigma}}{\partial v_{\|}} = 0,
  	\end{equation}
  	for the kinetic species $\sigma$ is solved together with the field equations (quasineutrality and Ampère's Law) in both codes. 
  	
  	A $\delta f$ splitting of the distribution function, $F=F_{M{\sigma}} + \delta f_{\sigma}(t)$, is used, with $F_{M{\sigma}}$ the Maxwellian, which is assumed as the equilibrium distribution function. With this splitting, an equation for the $\delta f_{\sigma}$ can be obtained:
  	
  	\begin{equation}
  	\frac{\partial \delta f_{\sigma}}{\partial t} = - \frac{\partial {\vec{R}}}{\partial t}  \nabla \delta f_{\sigma} - \frac{\partial v_{\|}}{\partial t} \frac{\partial \delta f_{\sigma}}{\partial v_{\|}} - \frac{\partial {\vec{R}}}{\partial t}  \nabla F_{\sigma} - \frac{\partial v_{\|}}{\partial t} \frac{\partial F_{\sigma}}{\partial v_{\|}} .
  	\label{deltaFEqnt}
  	\end{equation}
  	
  	EUTERPE uses a particle-in-cell scheme and the equation for $\delta f$ is solved by discretizing the distribution function using quasi-particles or markers. The equation is solved along the characteristic curves (defined by $\frac{\partial {\vec{R}}}{\partial t}$ and $\frac{\partial v_{\|}}{\partial t} $),  which are the equations of motion of the markers. 
  	PEST ($s,\theta,{\phi}$) magnetic coordinates \cite{Grim83} are used for the description of the fields, with $s={\Psi/\Psi_0}$, $\Psi$ the toroidal flux, $\Psi_0$ the toroidal flux at the {last closed flux surface}, ${\phi}$ the toroidal angle, and $\theta$ the poloidal angle. Cylindrical coordinates are used for the markers.   	
  	Several different formulations ($p_{\|}$, $v_{\|}$ o mixed variable scheme)  are presently implemented in the code \cite{Mishchenko14,Mishchenko14b}. In this work, the {$p_{\|}$ formulation is used and the equations solved are those used in references \cite{Riemann16,Sanchez20}, to which we refer the reader for details.  
 		
 	GENE-3D solves the GK Equation (\ref{deltaFEqnt}) in a fixed grid in the five-dimensional phase space (plus time), consisting of two velocity coordinates $v_{\parallel}$ (velocity parallel to the magnetic field) and $\mu$ (magnetic moment), and the three magnetic field-aligned coordinates $(x,y,z)$, with $x$ the radial coordinate, defined as $x=a\sqrt{s}$, $y$ the coordinate along the binormal direction and $z$ the coordinate along the field line.

 	In this work, the electrostatic equations are always solved in EUTERPE, both in adiabatic-electron and kinetic-electron simulations. Then, the gyrokinetic equation (\ref{deltaFEqnt}) is solved together with the quasineutrality equation. In the case of GENE-3D, the electromagnetic version of equation \ref{deltaFEqnt} is solved together with the quasineutrality, Ampère's law and an equation for the inductive electric field (Ohm's law) in the kinetic-electron simulations with GENE-3D presented in Section \ref{secCompLinearSimsW7X} while the electrostatic version of the equation \ref{deltaFEqnt} is solved in the simulations with adiabatic electrons.

EUTERPE can simulate {the entire confined plasma or a reduced volume covering a radial annulus with inner and outer radial boundaries other than the magnetic axis and the last closed flux surface} \cite{Jost01,Kornilov04}. 
For the {field solver}, natural boundary conditions are set at the inner boundary for the full {volume} simulations, while Dirichlet conditions are used {in the simulation considering an annulus}. Dirichlet conditions are always used  at the outer boundary. Density and temperature profiles, depending only on the radial coordinate, are used as input for both kinds of simulations.
GENE-3D can be used in either the flux tube, the full flux surface or the radially global domains \cite{Maurer19,Banon20,Wilms21}. In this work, only simulations in the  global domain will be used and Dirichlet conditions are used  at both the inner and the outer boundaries.

Both GENE-3D and EUTERPE use a real space representation of the fields (in their internal coordinates).
In EUTERPE, the electrostatic potential is Fourier transformed on each flux surface, and filtering the potential in Fourier space (with a variety of different filters) is possible. The code allows extracting a phase factor in linear simulations, thus allowing for a significant reduction in the computational resources \cite{Jost01}. 
Two different implementations of the quasi-neutrality equation and the field solver are available, one assuming a long wavelength approximation and another one using a Padé {approximant}, which allows resolving modes with arbitrarily large wavenumber (see \cite{Sanchez19} for details on these approximations). The Padé approximation \cite{Sanchez19} is used in this work, which allows resolving modes $k_{\perp}\rho > 1$. Keeping these modes in the simulation is required because in W7-X the spectrum of linearly unstable ITG modes shows peak growth rates for $k_{\perp}\rho \gg 1$ \cite{SanchezNF2021}.  

 The equilibrium magnetic field is obtained from a magneto-hydrodynamic equilibrium calculation carried out with the code VMEC \cite{Hirshman83}, and selected magnetic quantities are mapped for its use in the gyrokinetic codes with the intermediate programs VM2MAG and GVEC for EUTERPE and GENE-3D, respectively. 
 {In GENE-3D the resolution used in the equilibrium magnetic field is coupled to the resolution in the electrostatic potential, while in EUTERPE, both resolutions are independent. }
 
 Several tools are implemented in EUTERPE for the control of numerical noise and stabilization of the profiles in gradient-driven nonlinear simulations \cite{Sanchez20}. In this work, a heating source term is used to keep the temperature profile stable during the simulation and the weight smoothing is used for improving the signal to noise ratio.
 Particle and heat source terms as described in \cite{McMillan2009} are used in GENE-3D for sustaining the density and temperature profiles in nonlinear gradient-driven simulations. 
 For more details about the codes and the equations they solve, the reader is referred to \cite{Jost01,Kornilov04, Sanchez20,Slaby18} and \cite{Maurer19,Banon20,Wilms21}. 
% 

 %*******************************************
 \section{Comparison of the codes in linear simulations}\label{secCompLinearSims}
 %*******************************************

  %*******************************************
  \subsection{Linear simulations in LHD}\label{secCompLinearSimsLHD}
  %*******************************************
  
 In order to ease the comparison and to make the simulations less expensive computationally we use model density and temperature profiles with gradients well localized radially and with non realistic (because of the large  temperature) plasma parameters. We use the same model profiles from \cite{SanchezNF2021} for this comparison, which 
 are shown in Figure \ref{fig:profiles}. 
 The electron density is flat and the density and ion temperature are defined by the formula 
 \begin{eqnarray}
 X = {X_r} \rm{exp} \left[\frac{-\kappa_x}{1-\rm{sech}^2_{x} } \left( \tanh(\frac{r - r_0}{a\Delta_x}) - \rm{sech}^2_{X} \right)\right],
 \label{analProfilesnT}
 \end{eqnarray}
 where $X=\{n,T_i\}$ represents a radial profile of density ($n_e=n_i=n$) or ion temperature.

	 %*******************************************
	 \begin{figure}[!h]
	 	\centering
	 	\includegraphics[height=5.25cm, trim=5 0 0 0, clip, angle=0]{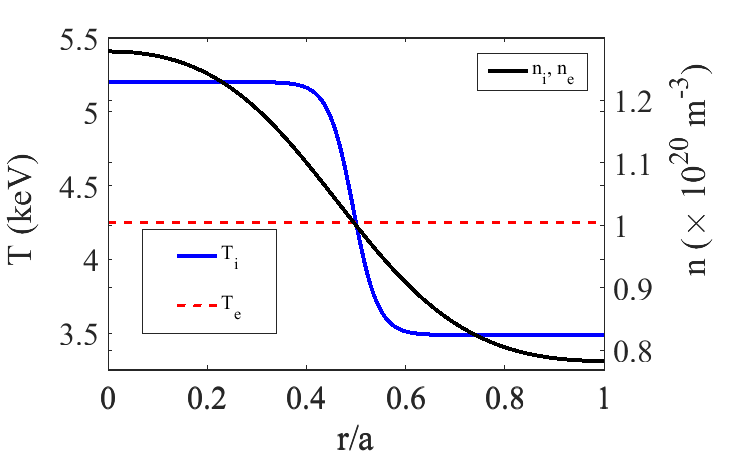}
	 	\includegraphics[height=5.25cm, trim=10 0 70 0, clip, angle=0]{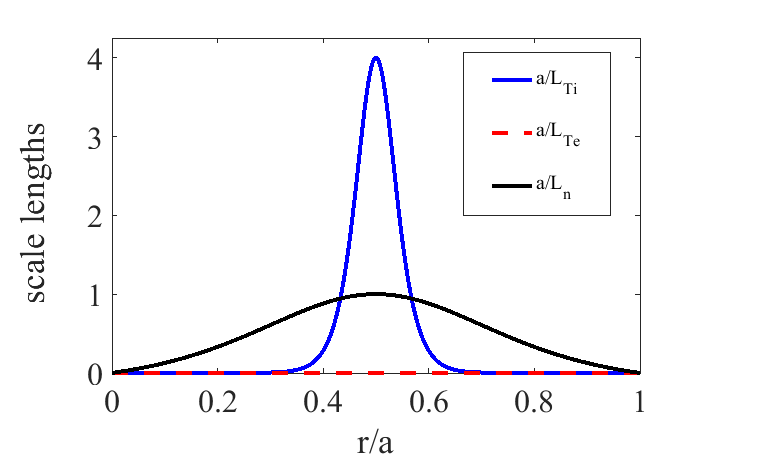}
	 	\caption{Density and temperature profiles used in the linear simulations in LHD (top) and the corresponding normalized scale lengths, $a/L_n$, $a/L_{Ti´}$ and $a/L_{Te}$ (bottom).}
	 	\label{fig:profiles}
	 \end{figure}
	 %*******************************************
 
 Some relevant parameters of these profiles are 
     $r_0=0.5~a$, $\Delta n = 0.3$,  $\Delta T_i=0.1$, $T_r=T_i(r/a=0.5) = T_e = 4250 ~eV$, $n_r=n_i = n_e = 10^{20} ~m^{-3}$, $a/L_{Ti}(r/a=0.5) = -4$, $ a/L_{Te}(r/a=0.5) = 0$, $a/L_{ni}(r/a=0.5) = -1$, $a/L_{ne}(r/a=0.5) = -1$, with $\rm{sech}^2_{X}=\cosh^{-2}(\frac{r_0}{a\Delta_X})$, and the scale length defined as $L_{X}=|\frac{1}{X}\frac{dX}{ds}|^{-1}$.  We normalize the frequencies and growth rates with $v_{th}/a$, with $v_{th} = \sqrt{T_i(r/a=0.5)/m_i}=6.3612 \cdot 10^{5}~\rm{m/s}$ the reference thermal velocity, $m_i$ the proton mass, and $a=0.598$  the minor radius (\texttt{Aminor\_p} obtained from VMEC), $r/a=\sqrt{s}$ is the normalized radius,    with $s$ the normalized toroidal flux and $v_{th}/a = 1.0638\times 10^{6}~\rm{s}^{-1}$.
 
 With these density and temperature profiles we carry out linear simulations with adiabatic electrons using both codes and perform a thorough comparison of results. In these global linear simulations we identify the most unstable mode, for which we will characterize its toroidal and poloidal mode numbers, its mode structure and spatial location and will extract its growth rate and frequency. From previous works, we know that the spectrum of linearly unstable ITG modes in LHD is compact, with maximum growth rate for normalized mode numbers $k_{\perp}\rho<1$, similar to that in a tokamak \cite{SanchezNF2021,Regana2021}. Then, the simulations can be run with moderate resolutions and are relatively cheap in terms of computational resources. 
The simulation carried out with EUTERPE is global in radius while it is restricted to an annulus  ($0.2 < r/a<0.8$) in the case of GENE-3D. The resolutions used in the EUTERPE simulation are $n_s \times n_{\theta} \times n_{\phi} = 64\times256\times64$. In GENE-3D, resolutions are $n_x \times n_y \times n_z \times n_{\parallel} \times n_{\mu} = 144 \times 64 \times 128 \times 48 \times 20$. In the following sections, we study and compare in detail the properties of the unstable modes found with both codes.

%*******************************************
\subsection{Mode structure and spatial location}\label{subSecModeStruct}
%*******************************************

Figure \ref{fig:EUTGENE_LHDPotRZ} shows the amplitude of electrostatic potential at the plane with toroidal angle $\phi=0$ in two simulations with EUTERPE and GENE-3D.
		
		%*******************************************
		\begin{figure}[!h]
			\centering
			\includegraphics[height=7.5cm,  trim=20 0 45 50, clip, angle=0]{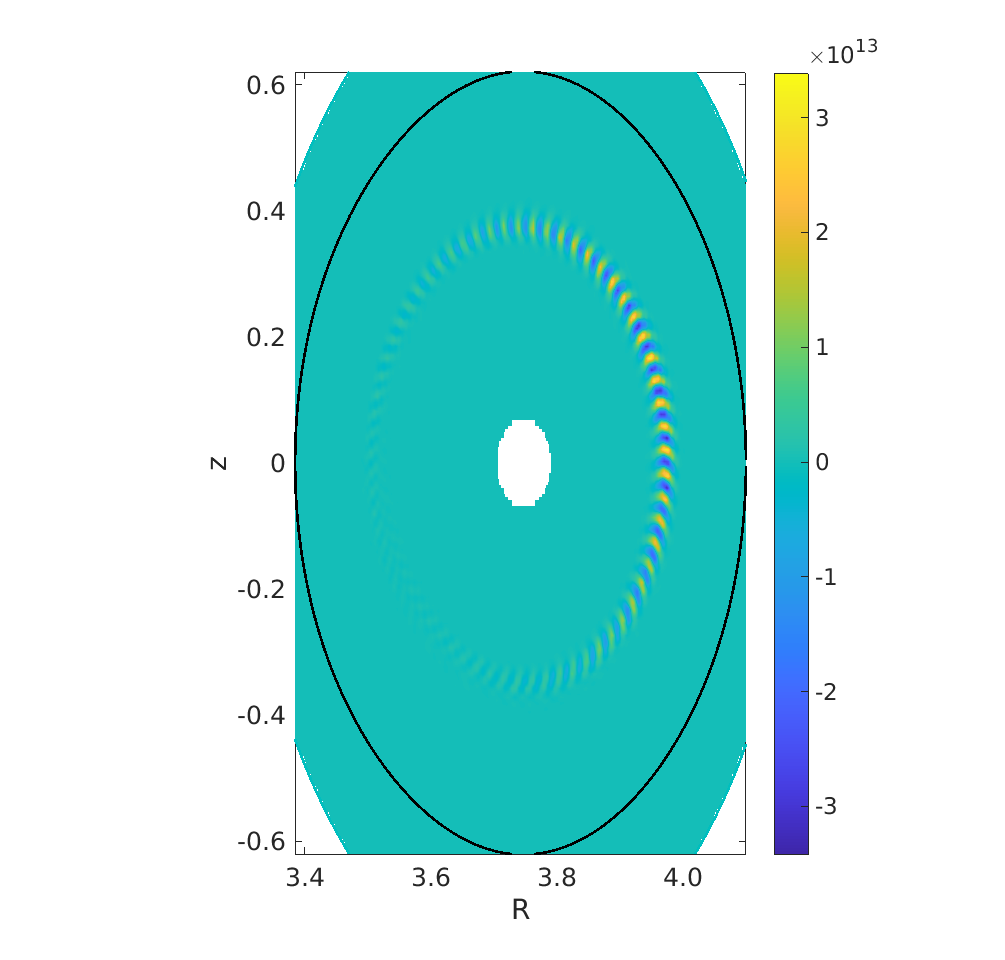}
			\includegraphics[height=7.5cm, trim=120 0 50 45, clip, angle=0]{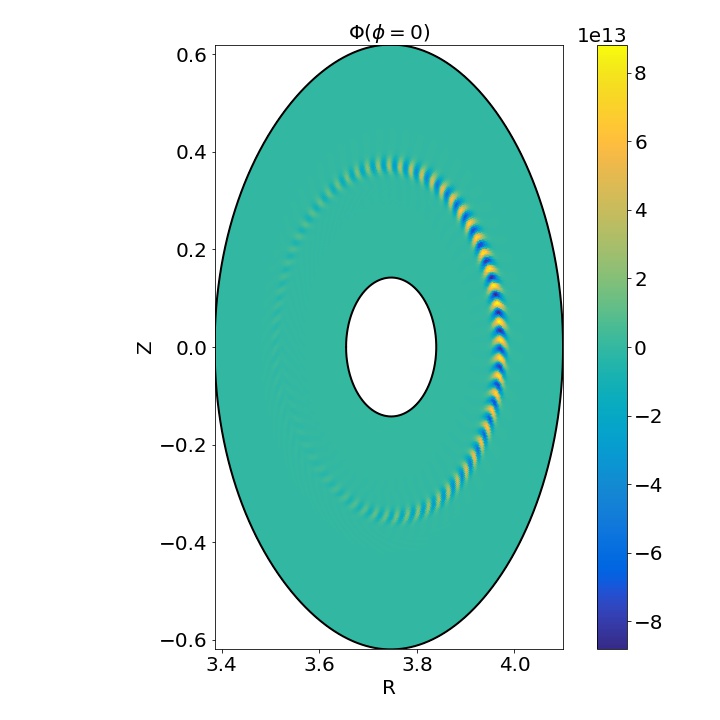}
			\caption{Amplitude of the electrostatic potential (normalized to its maximum) in a  plane at toroidal angle ($\phi=0$) for two simulations carried out with EUTERPE (left) and GENE-3D (right) in the LHD standard configuration using the profiles from Figure \ref{fig:profiles}. Note the different radial limits in both panels. The EUTERPE simulation is full global while the GENE-3D one is carried out in a radial annulus $0.2 < r/a<0.8$. A line is shown for $r/a=0.8$ in the EUTERPE case (left).}
			\label{fig:EUTGENE_LHDPotRZ}
		\end{figure}
		%*******************************************

The excellent agreement in the mode found by both codes is already apparent in this figure and will be analyzed in more detail next. Figure \ref{fig:EUTGENE_LHDPotrtheta} shows another view of the mode in the plane $(r/a, \theta)$ at $\phi=0$ in both simulations. 

The mode propagates in the ion diamagnetic direction as corresponds to an ion-temperature-gradient-driven mode. 
		%*******************************************
		\begin{figure}[!h]
			\centering	
			\includegraphics[width=6.5cm, trim=20 20 0 0, clip, angle=0]{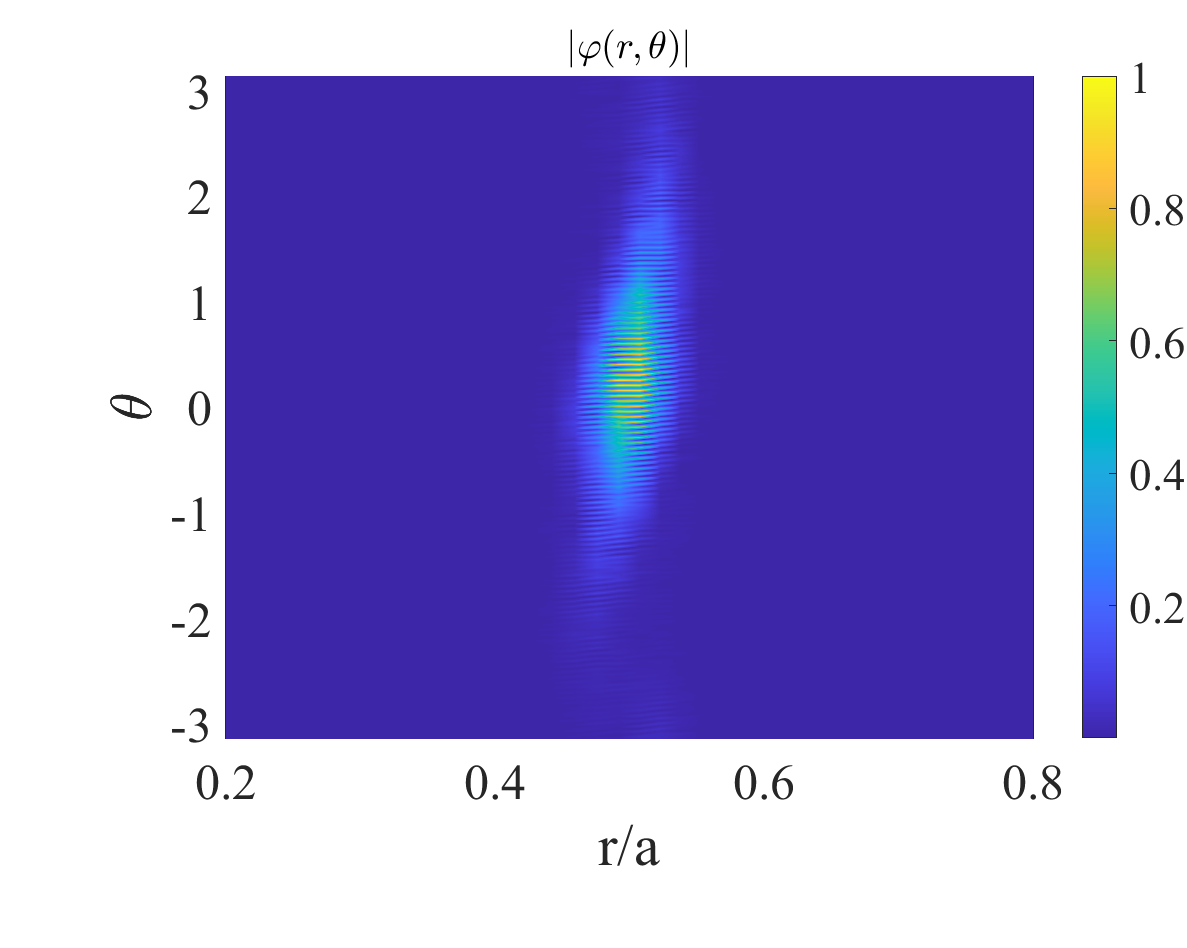}
			\includegraphics[width=6.5cm, trim=20 20 0 0, clip, angle=0]{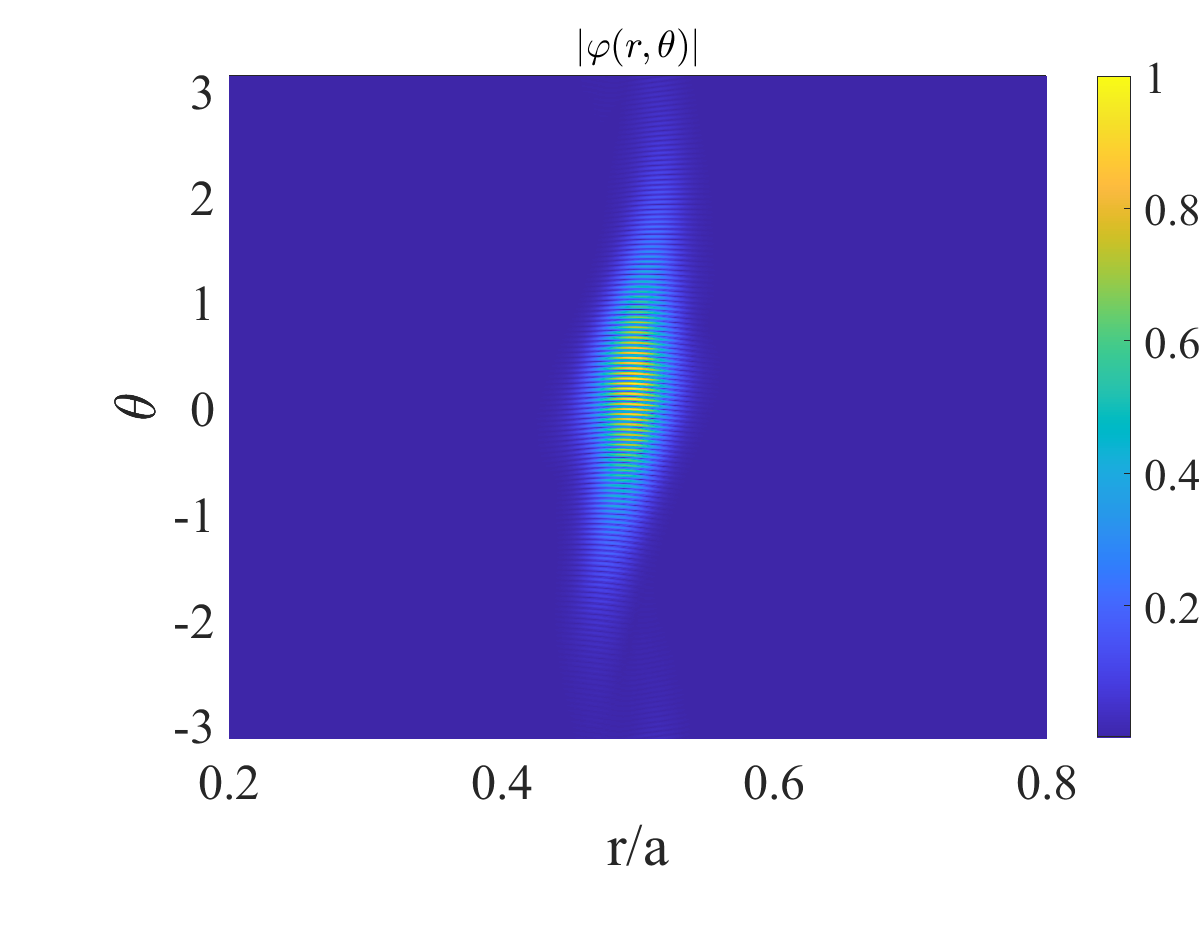}
			\caption{Amplitude of the electrostatic potential (normalized to its maximum value) vs radial ($r/a$) and poloidal ($\theta$) coordinates for the two simulations carried out with EUTERPE (left) and GENE3D (right) in LHD using the profiles from Figure \ref{fig:profiles}}
			\label{fig:EUTGENE_LHDPotrtheta}
		\end{figure}
		%*******************************************
The largest amplitude of potential is found around $(r/a=0.5$) showing some dependence of the radial location of maximum amplitude perturbation with the poloidal angle, $\theta$, with good agreement between the codes. 

%*******************************************
\subsection{Fourier spectrum}\label{subSecFourierMode}

%*******************************************
Now we will analyze the Fourier spectrum of the perturbed potential in both simulations. The poloidal and toroidal spectra at the planes  $\phi=0$ and $\theta=0$, respectively, are shown in Figure \ref{fig:EutLHDLinModesm}. Very good agreement between the codes is also found with respect to the Fourier spectrum of the electrostatic potential. Peak spectral amplitude  is found around $n=30$ and $m=64$ by both codes. 
%
		%*******************************************
		\begin{figure}[!h]
			\centering
			
			\includegraphics[width=6.5cm, trim=20 10 0 0, clip, angle=0]{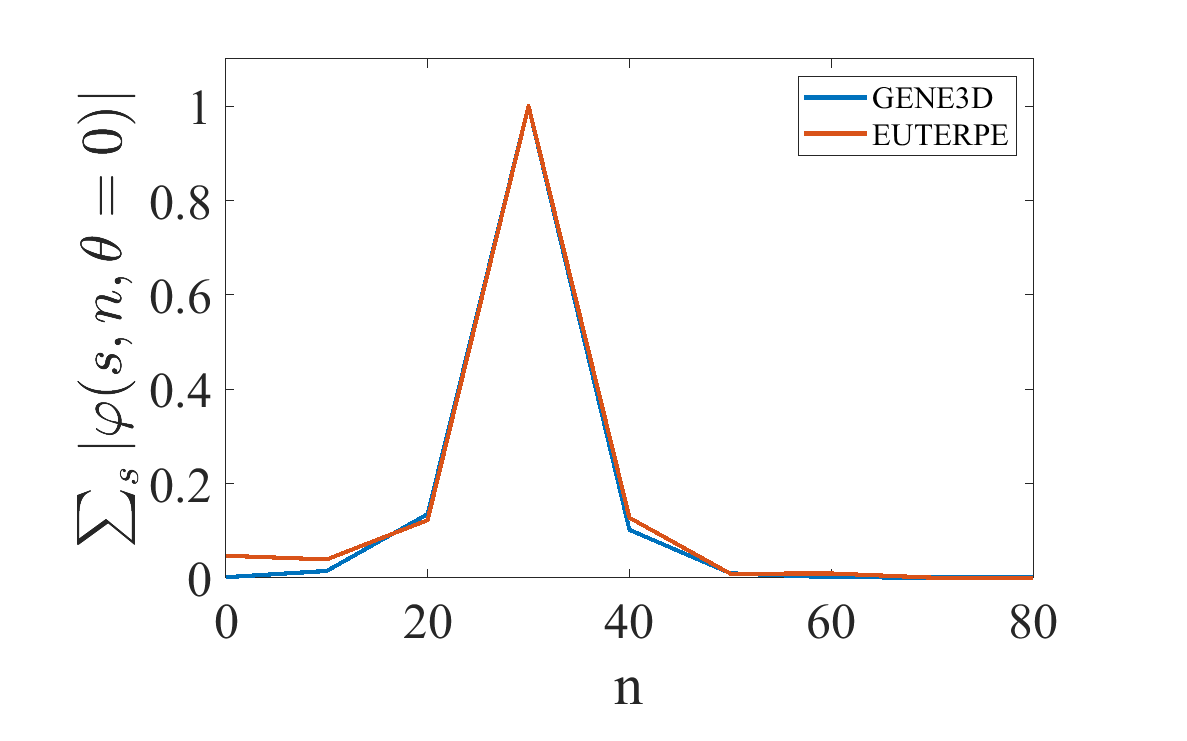}
			\includegraphics[width=6.5cm, trim=20 10 0 0, clip, angle=0]{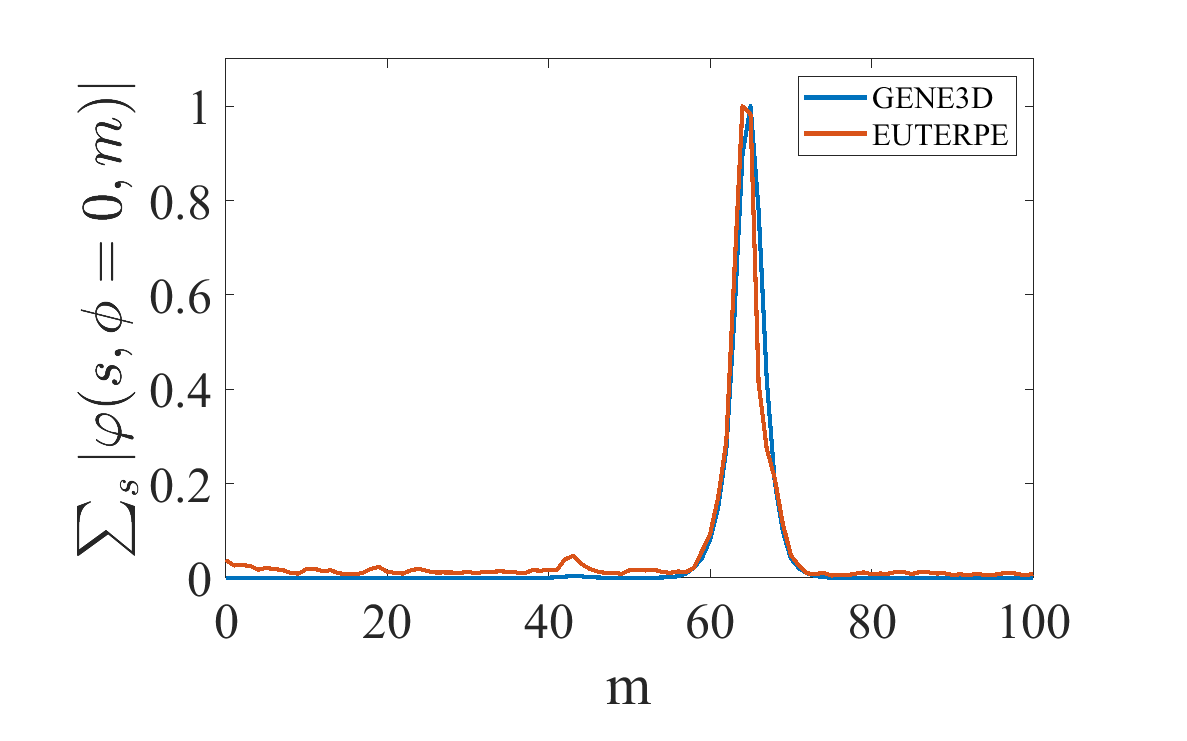}
			\caption{Spectral amplitude of (normalized) toroidal (left) and poloidal (right) spectra for at the planes  $\phi=0$  and $\theta=0$, respectively, for the two simulations carried out with EUTERPE and GENE-3D  in LHD using the profiles from Figure \ref{fig:profiles}}
			\label{fig:EutLHDLinModesm}
		\end{figure}
		%*******************************************
%
Figure \ref{fig:EUTGENE_LHDPotr} shows the integrated spectral amplitude of the potential,  $\sum_n|\varphi(s, n, \theta=0)|$ and $\sum_m|\varphi(s,\phi=0,m)|$, vs radial coordinate at the planes $\theta=0$ and $\phi=0$ for both simulations. The agreement in the radial location of the maximum amplitude of perturbed potential is even clearer in this figure. A strong localization of this potential perturbation in the radial coordinate, which is related to the strong $T_i$ gradient at middle radius shown in the temperature profiles in Figure \ref{fig:profiles}, is found in both codes.
%*******************************************
\begin{figure}[!h]
	\centering	
	\includegraphics[width=6.5cm, trim=20 10 0 0, clip, angle=0]{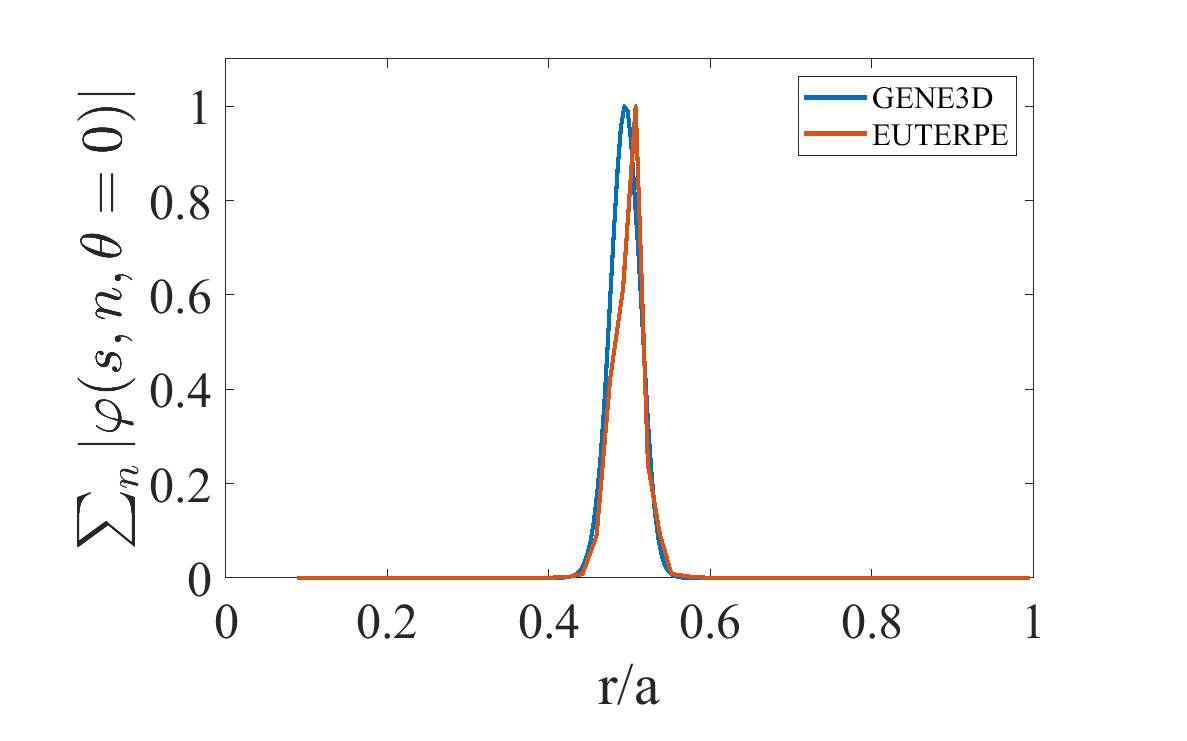}
	\includegraphics[width=6.5cm, trim=20 10 0 0, clip, angle=0]{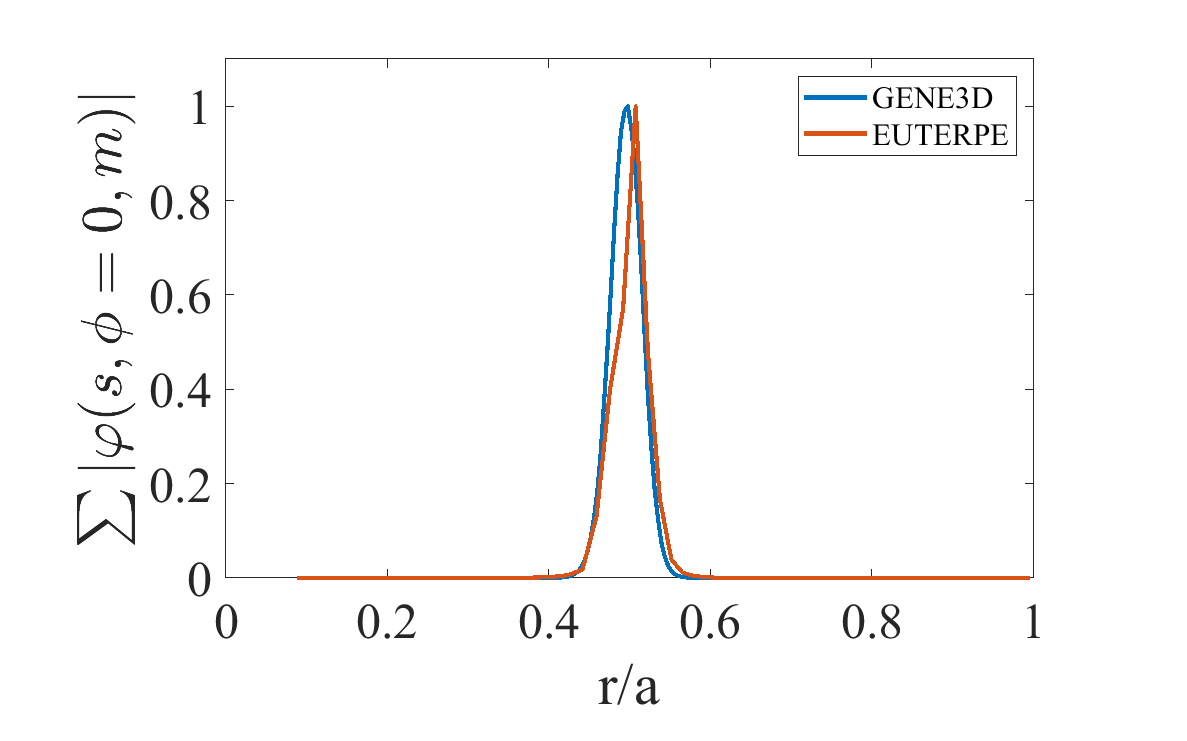}
	\caption{Radially integrated spectral amplitude of the electrostatic potential (normalized) vs radial ($r/a$) coordinate at the planes $\theta=0$ ($\sum_n|\varphi(s, n, \theta=0)|$) and $\phi=0$ ($\sum_m|\varphi(s,\phi=0,m)|$) for the  simulations carried out with EUTERPE and GENE-3D  in LHD using the profiles from Figure \ref{fig:profiles}}
	\label{fig:EUTGENE_LHDPotr}
\end{figure}
%*******************************************		

For a more complete and quantitative description of the most unstable mode we extract the linear growth rate and frequency of the mode. For this, we make a fit of the time signal of the electrostatic potential for the most unstable mode to an exponentially growing  oscillation with the form $\varphi(t)= \varphi_0 e^{(i\omega  + \gamma)t}$. The most unstable mode is identified as that with the largest amplitude at the end of the simulation (or the time window chosen for the analysis). Exactly the same toroidal mode number, $n=30$ is obtained in both codes, while there is a slight mismatch in the poloidal mode number; $m=64$ is obtained with EUTERPE while $m=65$ is obtained in GENE-3D. 
Note that the poloidal mode number, $m$,  is not extracted exactly the same way in both codes. In EUTERPE, the toroidal and poloidal spectra at planes $\theta=0$ and $\phi=0$ are registered for many time instants along the simulation and from these, the most unstable toroidal and poloidal mode, respectively, are identified. The real and imaginary parts of the potential corresponding to the most unstable toroidal/poloidal mode are fit to the model and the growth rate $\gamma$ and frequency $\omega$ extracted. In this way, independent fits are carried out for the toroidal and poloidal spectra, and the toroidal and poloidal mode numbers of the most unstable mode are extracted independently. As expected, modes close to resonant ($n \approx \iota m$) are obtained. 
To extract the toroidal and poloidal modes number in GENE-3D, a transformation of coordinates between the field-aligned  coordinates ($x,y,z$) to PEST coordinates ($x,\phi,\theta$) is carried out, and then the Fourier transform of the later is taken for the latest time step of the linear simulation. The growth rate, $\gamma=0.21 ~v_{th}/a$ and frequency, $\omega=0.39~v_{th}/a$, of the most unstable mode obtained with EUTERPE are in good agreement with those obtained with GENE-3D, $\gamma=0.20 ~v_{th}/a$ and $\omega=0.40~v_{th}/a$,  with $v_{th}$ the ion thermal velocity and $a$ the minor radius. 
In terms of normalized wavenumbers $m=64,65$ correspond to {$k_{\theta}\rho_i=0.56,0.57$, with  $k_{\theta}=m/r$}, which are spatial scales  similar to those expected for the most unstable ITG modes in a tokamak.

 %*******************************************
 \subsection{Linear simulations in W7-X}\label{secCompLinearSimsW7X}
 %*******************************************
 
 In W7-X, the comparison  of linear simulations is not as easy as in LHD. From previous works, \cite{SanchezNF2021,Regana2021} we know that the spectrum of ITG unstable modes is much wider in W7-X than that in LHD, with peak growth rates for $k_{\perp}\rho_i \gg 1$ in W7-X. This implies that a huge resolution is required to properly resolve the most unstable mode in W7-X, which makes these global simulations more expensive computationally that simulations in LHD. If moderate resolution is used in the simulations, the modes that appear as the most unstable in a simulation largely depend on the response of the code in Fourier space, the details of the numerical implementation, the coordinates used, etc. Because of this, single point comparison of simulations as we performed in LHD are not suited in this case, except we use a  huge resolution that can capture and properly resolve modes up to {$k_y\rho_i \gg 5$} or more (see \cite{SanchezNF2021,Regana2021}). We cannot expect perfect agreement between the codes on the most unstable mode if we use smaller resolutions. To overcome these difficulties, we compare the codes in a set of simulations with increasing resolutions in the angles $(\theta, \phi)$ over the flux surface in EUTERPE and the resolution in $y$ coordinate in GENE-3D. This way we can plot the curves for $\gamma$ and $\omega$ (of the most unstable mode) vs toroidal mode number for each code and compare them. We will compare the codes in both adiabatic-electron and kinetic-electron simulations.
 %
 %*******************************************
 \subsubsection{Adiabatic-electron simulations.}\label{subsecW7XLinAE}
 %*******************************************
 %
 
 We have run a set of adiabatic-electron simulations with EUTERPE with resolutions $n_{\theta} =256, 384, 512, 768, 1024$ using the same profiles from Figure \ref{fig:profiles}. The resolution in the toroidal angle is increased accordingly when the resolution in the poloidal direction is increased, as to capture near-resonant modes.  With GENE-3D, resolutions in the binormal direction $n_y= 64, 128, 256, 512$ were used. The resolution in the parallel direction, $n_z=128$ is kept constant, as it is the minimum resolution in $z$ required to properly resolve the equilibrium quantities and is considered to be sufficient for resolving the fields in all the cases.
 Figure \ref{fig:compEUTGENE_W7XAEwScan} shows the growth rate and frequency of the most unstable mode vs the toroidal mode number, $n$, for these two sets of  simulations.

%*******************************************
\begin{figure}[!h]
	\centering
	
	\includegraphics[width=7.5cm, trim=20 10 50 0, clip, angle=0]{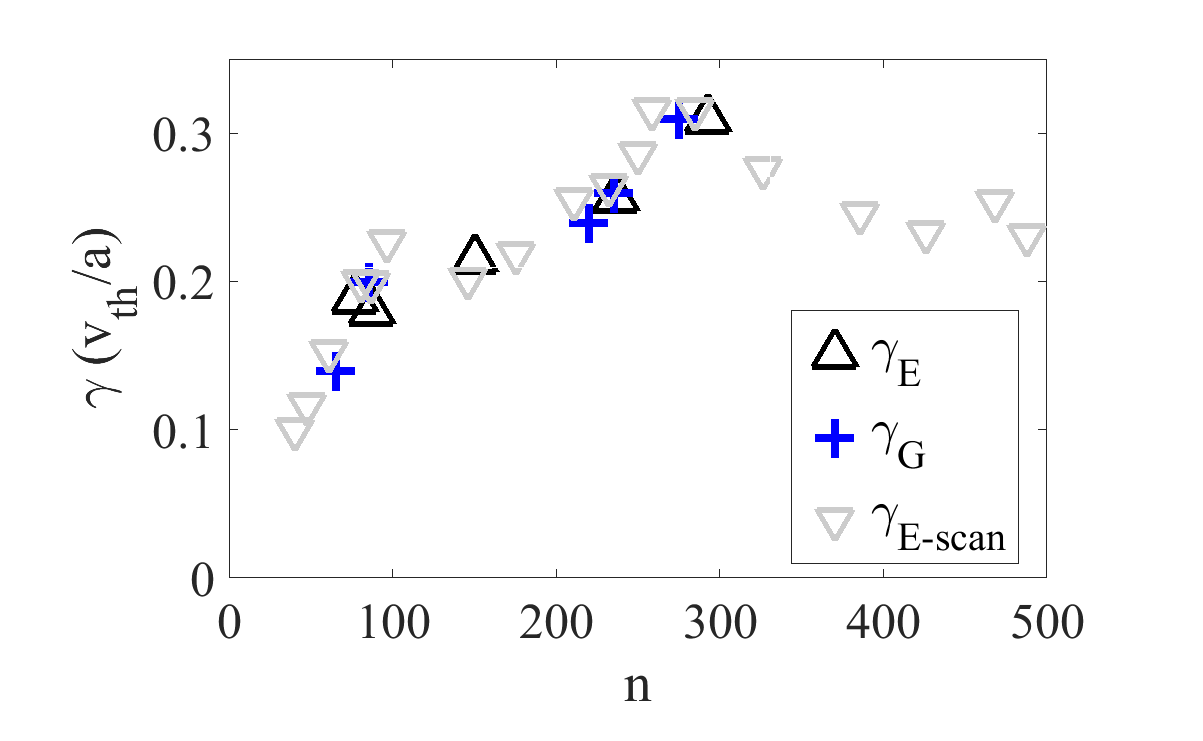}
	\includegraphics[width=7.5cm, trim=20 10 50 0, clip, angle=0]{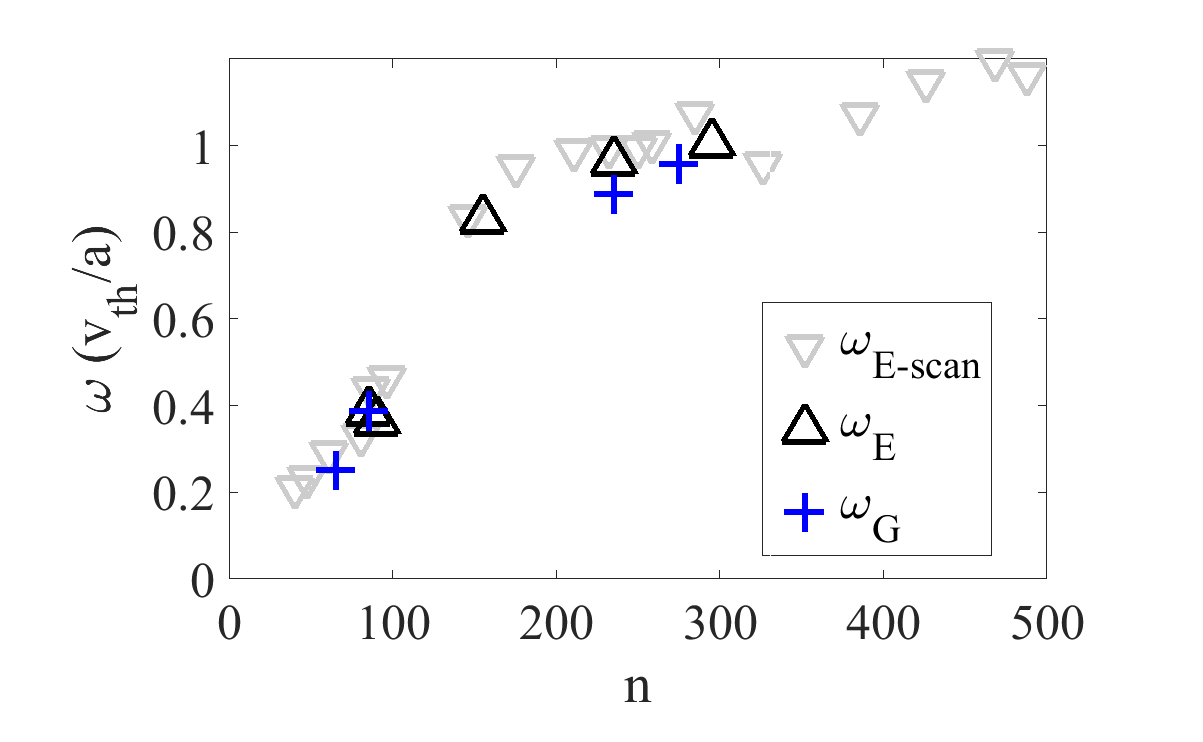}
	\caption{Growth rate (left) and frequency (right) of the most unstable mode for the simulations carried out with EUTERPE (subscript "E") and GENE-3D (subscript "G") in W7-X using the profiles from Figure \ref{fig:profiles} and adiabatic electrons. Results of a scan using a phase factor and a narrow filter (subscript "E-scan") are also shown (see text for details).}
	\label{fig:compEUTGENE_W7XAEwScan}
\end{figure}
%*******************************************
 It is clear from the figure that there is a good agreement between the codes on the $\gamma-n$ and $\omega-n$ curves, while the agreement in a point-to-point basis is not so good, which can be explained by the different numerical implementations in the codes, that imply different response, as discussed before. The largest resolution simulations with both codes can resolve modes up to $n=445$ %$m=512$ 
 in the EUTERPE case and $n=1270$ in the GENE-3D one, however it is found that those small-scale modes are not dominant, and the most unstable ones are below $300$ ($n_{max}=295$ in EUTERPE and $n_{max}=275$ in GENE-3D), that would correspond to $k_{\theta} \rho_i \approx 3$.  
The modes propagate in the ion diamagnetic direction, as expected for ITG modes.

 EUTERPE implements the feature of phase factor extraction \cite{SanchezNF2021} that allows reducing the computational cost for linear simulations, by focusing the Fourier filter at specific region of the spectrum. Using this functionality we have run a set of simulations with smaller resolution {($n_s\times n_{\theta}\times n_{\phi}=64\times 128\times 32$)} and changing the center of the Fourier filter, thus scanning the spectrum at a moderate computational cost. Figure \ref{fig:compEUTGENE_W7XAEwScan} shows the results obtained with this scan (labeled "E-scan") compared with the results of larger resolution simulations. 	
 A good agreement is found between the smaller resolution set of simulations and those with larger resolutions with GENE-3D and EUTERPE. In addition, a drop in the growth rate for $n>300$, which cannot be resolved with large resolution simulations that are dominated by the most unstable mode in the full Fourier domain resolved, is confirmed by the scan results.

 %*******************************************
 \subsubsection{Kinetic-electron simulations.}\label{subsecW7XLinKE}
 %*******************************************
 In addition to the adiabatic-electron simulations just described, 
 we ran simulations using the same profiles as in the last section  (see Figure \ref{fig:profiles}) and including both the ions and the electrons as kinetic species. The simulations are collisionless and the results obtained for the growth rate and frequency of the most unstable mode are shown in Figure %s \ref{fig:compEUTGENE_W7XKE} and 
  \ref{fig:compEUTGENE_W7XKE2}.

 A set of simulations with kinetic electrons is run with GENE-3D with resolutions $n_y= 64, 128,  256$. 
 The simulations are electromagnetic with $\beta=10^{-4}$, which ensures a negligible influence of electromagnetic effects.
 The simulations with $n_y=128$ and $n_y=256$ provide exactly the same growth rate and frequency for the most unstable mode.
 A set of simulations including kinetic electrons has also been run with EUTERPE. In this case the phase factor extraction functionality is used to reduce computational resources and the resolutions used are $n_{\theta} \times n_{\phi}=64 \times 32$. 
 Reasonable agreement is found for the most unstable mode. In addition to results for the kinetic-electron simulations, the results from adiabatic-electron simulations presented in Section \ref{subsecW7XLinAE} are also shown in Figure \ref{fig:compEUTGENE_W7XKE2} for comparison.
 %
		 %*******************************************
		 \begin{figure}[!h]
		 	\centering
		 	
		 	\includegraphics[width=7.5cm, trim=20 10 50 0, clip, angle=0]{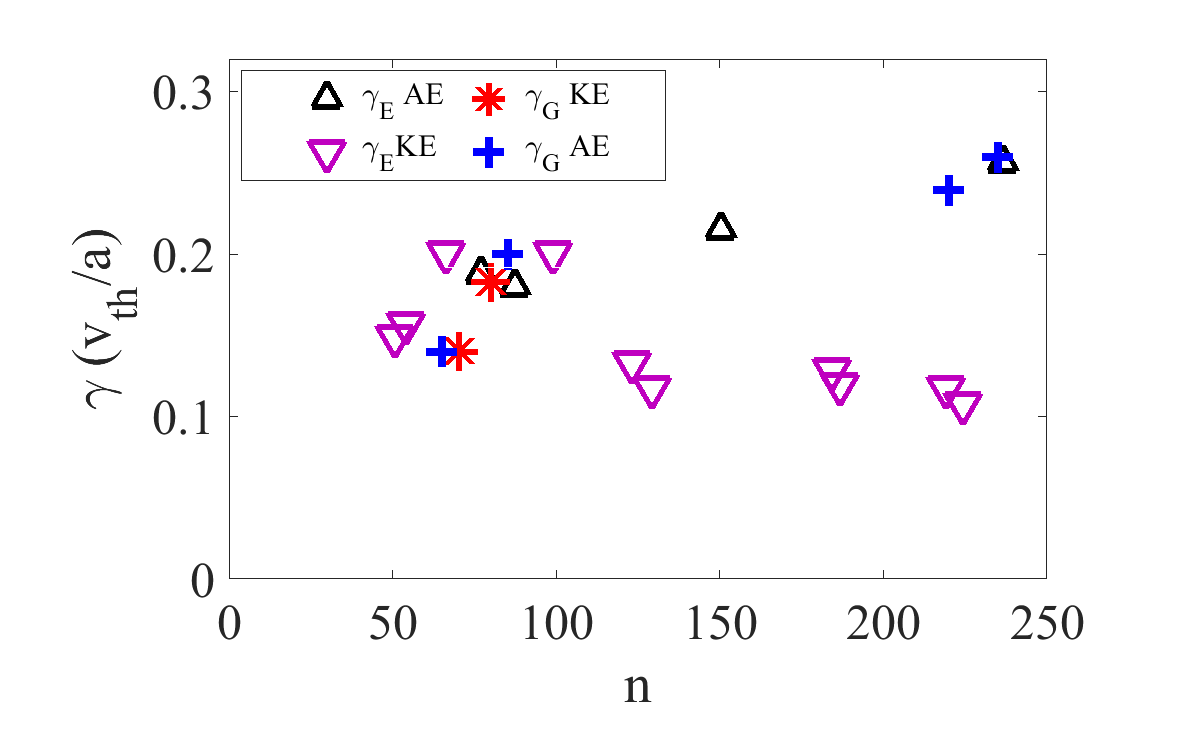}
		 	\includegraphics[width=7.5cm, trim=20 10 50 0, clip, angle=0]{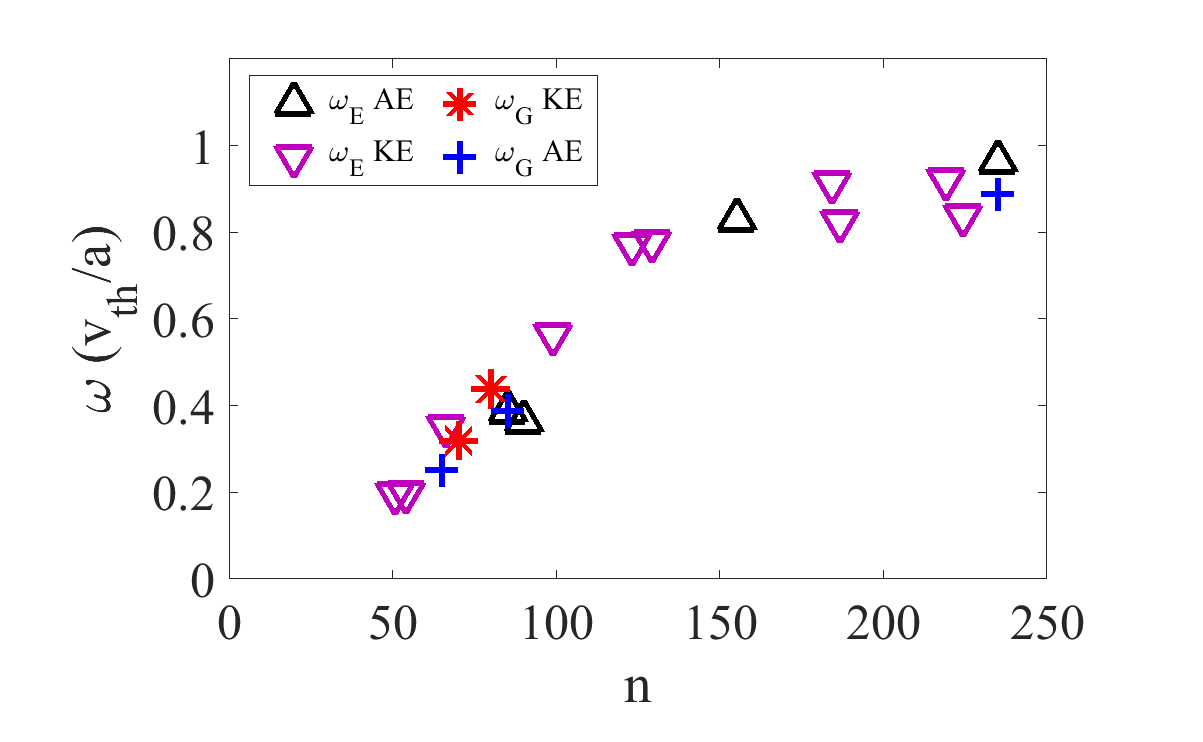}
		 	\caption{Growth rate (left) and frequency (right) of the most unstable mode for a set of simulations carried out with EUTERPE (subscript "E") and GENE-3D (subscript "G") in W7X using the profiles from Figure \ref{fig:profiles} and kinetic ions and electrons.  Results of the adiabatic-electron simulations are also shown for comparison.}
		 	\label{fig:compEUTGENE_W7XKE2}
		 \end{figure}
		 %*******************************************
A clear stabilizing effect is observed for modes with  $n>100$ in the simulations including kinetic electrons, while the growth rate for modes $n\sim 100$ is not much affected. The frequencies of all modes are insensitive to the influence of kinetic electrons.
%*******************************************
%*******************************************
\section{Comparison of nonlinear simulations}\label{secNonLin}
%*******************************************
%*******************************************
%*******************************************
%
\subsection{Nonlinear simulations in LHD}\label{secNonLinLHD}
%
%*******************************************
Using the same density and temperature profiles from Figure \ref{fig:profiles} we run nonlinear simulations with both codes and compare the turbulent heat flux extracted in the saturated nonlinear phase. Some simulations are restricted in radius in order to reduce the computational cost. For the comparison, we target 
%the volume integrated heat flux and also 
the ion heat flux.
% 
%*******************************************
\begin{figure}[!h]
	\centering
	\includegraphics[width=7.5cm, angle=0]{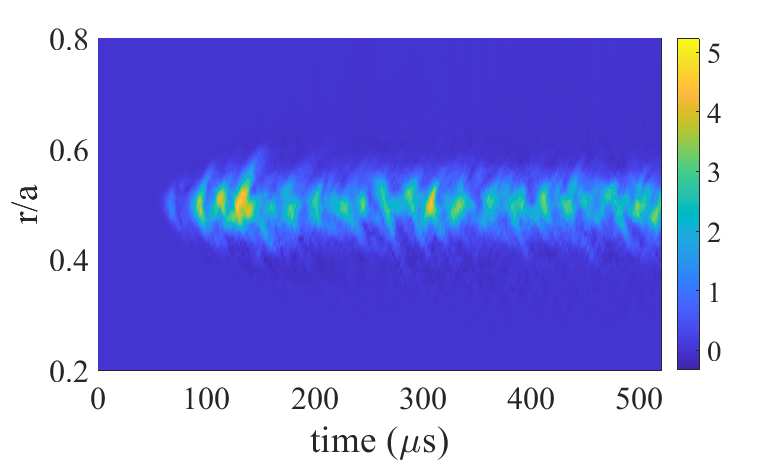}\\
	\includegraphics[width=7.5cm, angle=0]{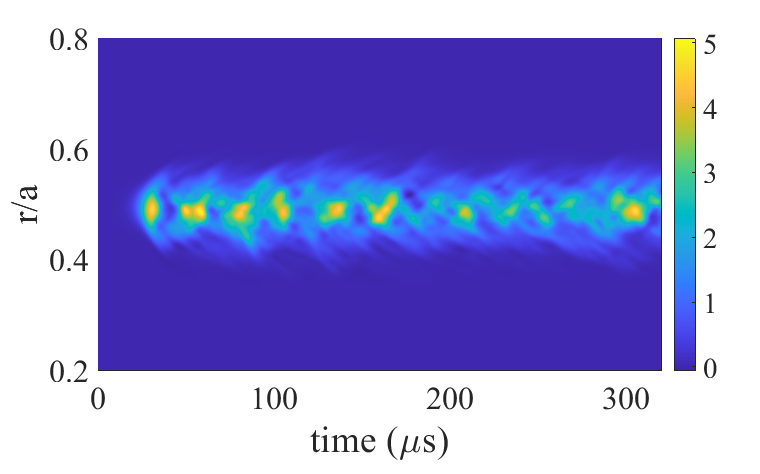}
	\caption{Turbulent ion heat flux versus radial position and time in two simulations carried out in the LHD configuration with EUTERPE (top) and GENE-3D (bottom) using the same kinetic profiles from Figure \ref{fig:profiles} and the same value of the heating source, $\gamma_h=\gamma_{max}/10$.}
	\label{fig:LHDHeatFluxVsRt}
\end{figure} 
%*******************************************

 In nonlinear simulations, the density and temperature profiles decay after the nonlinear saturation, which needs to be compensated by sources. We use an ad-hoc source term similar to that defined in \cite{McMillan2009}  to keep the profiles stable and reach a quasi steady state in which physically meaningful quantities can be extracted. Note that this source is not realistic and is used with the purpose of compensating the flattering produced by turbulent transport and sustaining the profiles. 
Figure \ref{fig:LHDHeatFluxVsRt} shows the turbulent ion heat flux versus radial position and time for two simulations carried out in the LHD configuration with EUTERPE and GENE-3D using the same heating source, $\gamma_h=\gamma_{max}/10$, with $\gamma_{max}$ the maximum growth rate in a linear simulation carried out with the same kinetic profiles. The heat flux is concentrated in the radial region where the temperature profile has a strong gradient (see Figure \ref{fig:profiles}) and the linear instability develops.

For a more quantitative comparison, in Figure \ref{fig:LHDHeatFluxVst}, we show a comparison of the turbulent ion heat flux vs time at radial position $r/a=0.5$ for the same two simulations.  %In Figure  \ref{fig:heatFluxvsrLHD} we will compare the radial profiles of this heat fluxe.
%*******************************************
\begin{figure}[!h]
	\centering
	\includegraphics[width=7.5cm, angle=0]{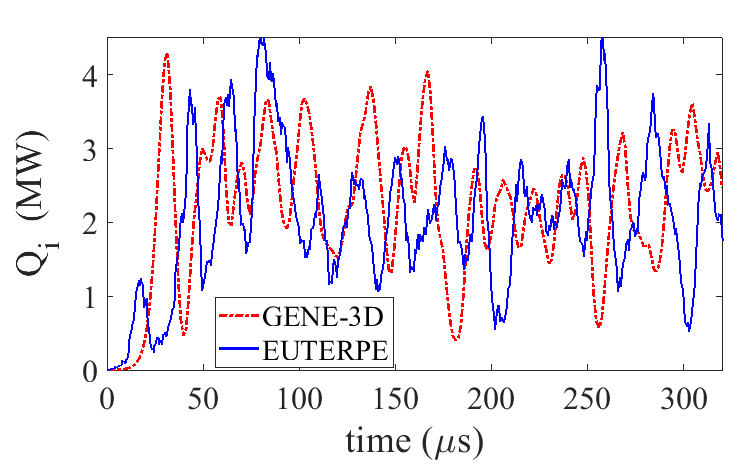}
	\caption{Surface-integrated ion heat flux at radial position $r/a=0.5$ in two simulations carried out in the LHD configuration with EUTERPE and GENE-3D using the same kinetic profiles from Figure \ref{fig:profiles} and the same value of the heating source, $\gamma_h=\gamma_{max}/10$.}
	\label{fig:LHDHeatFluxVst}
\end{figure}
%*******************************************
It can be appreciated in this figure that a stable value of the heat flux is reached after the saturation and also that there is a good agreement between the two codes in this heat flux level. Note that the time required to reach the saturation depends strongly on the initialization and numerical details of each code. In this case, the time trace corresponding to EUTERPE has been shifted $50~ \mu s$ to the left in Figure \ref{fig:LHDHeatFluxVst} for a better matching of the two time traces.  

A strong dependence with the strength of the heating source has been observed in the turbulent heat flux for these kinetic profiles, as it will be shown below. We have carried out a series of simulations with both codes using different values of the heating source (see \cite{McMillan2009,Sanchez20}). The values $\gamma_h \approx \gamma_{max}/2, \gamma_{max}/10, ~\rm{and}~ \gamma_{max}/20$ have been used. 
 The results for the surface-integrated turbulent heat flux versus the radial position are shown in Figure \ref{fig:heatFluxvsrLHD}. The value of heat flux at each radial position is obtained from a time average of the instantaneous heat flux in a long time after the nonlinear saturation in which a quasi-steady state is reached ($t>100~\mu s$ in Figure \ref{fig:LHDHeatFluxVsRt}).

%
		%*******************************************
		\begin{figure}[!h]
			\centering
			\includegraphics[width=7.5cm, trim=20 0 50 0, clip, angle=0]{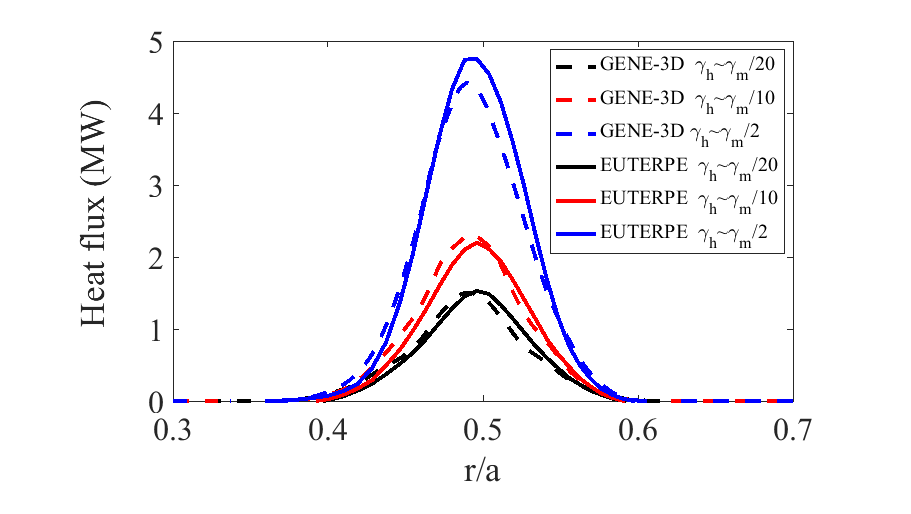}
			\caption{Turbulent (surface integrated) ion heat flux vs radial position in simulations carried out with EUTERPE and GENE-3D in the standard configuration of  LHD using the profiles from Figure \ref{fig:profiles}.}
			\label{fig:heatFluxvsrLHD}
		\end{figure}
		%*******************************************
The agreement between the codes on the heat flux for all the values of the heating source is very good. A clear dependence of the turbulent heat flux level with the strength of the source term is observed, which is in agreement with previous results, as discussed in \cite{McMillan2009, Sanchez20}. This strong dependence of the heat flux  with the strength of the source indicates that these kinetic profiles, with a very localized instability source, are not the best-suited for the comparison of heat fluxes. This set of simulations served us well to illustrate the dependence with heating source and to show the good agreement between the codes for different settings, however.
%
%*******************************************
%
\subsection{Nonlinear simulations in W7-X}\label{secNonLinW7X}
%
%*******************************************
%
The ideal profiles from Figure \ref{fig:profiles}, with very localized gradients are very different from real experimental profiles for which the region of instability is usually wider. Furthermore, keeping these profiles during the simulation and after the turbulence saturation using heating sources can result  difficult, even for a large value of the parameter $\gamma_h$, as discused before. As the simulations evolves, the temperature gradient decays at the radial regions of more turbulence activity and consequently the heat flux is reduced.

{In addition, in W7-X, the spectrum of linearly unstable ITG modes is much wider than that of LHD, which exacerbates the evolution of profiles and makes the comparison between the codes more difficult}. To overcome these difficulties, simulations for the W7-X configuration have been run with profiles having an instability less localized radially than that of profiles in Figure \ref{fig:profiles}. We have used profiles similar to those in \cite{McMillan2009}, which are shown in Figure \ref{fig:profilesMcMillan}.
		%*******************************************
		\begin{figure}[!h]
			\centering
			\includegraphics[height=4.5cm,  trim=30 10 50 0, clip, angle=0]{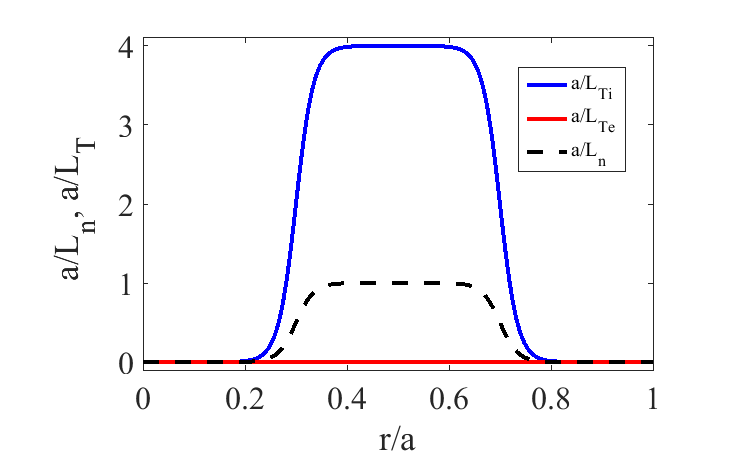}
			\includegraphics[height=4.5cm,  trim=30 10 20 0, clip, angle=0]{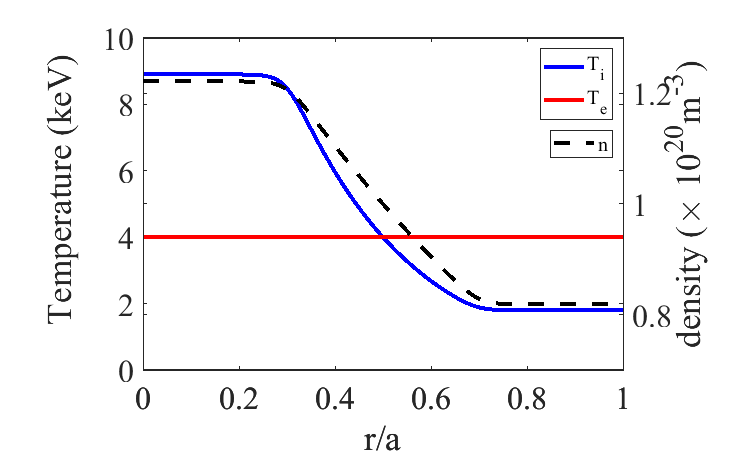}
			\caption{Density and temperature profiles used in the nonlinear simulations in W7-X from Section \ref{secNonLinW7X}.}
			\label{fig:profilesMcMillan}
		\end{figure}
		%*******************************************
Using these profiles we have run linear and nonlinear simulations with both GENE-3D and EUTERPE. Note that we need the linear simulations to set an appropriate value of the heating source that is significantly smaller than the maximum linear growth rate (see \cite{McMillan2009,Sanchez20}),  thus ensuring that the source does not introduce a too-large  perturbation.
		%*******************************************
\begin{figure}[!h]
	\centering
	\includegraphics[width=7.5cm,  trim=30 10 50 0, clip, angle=0]{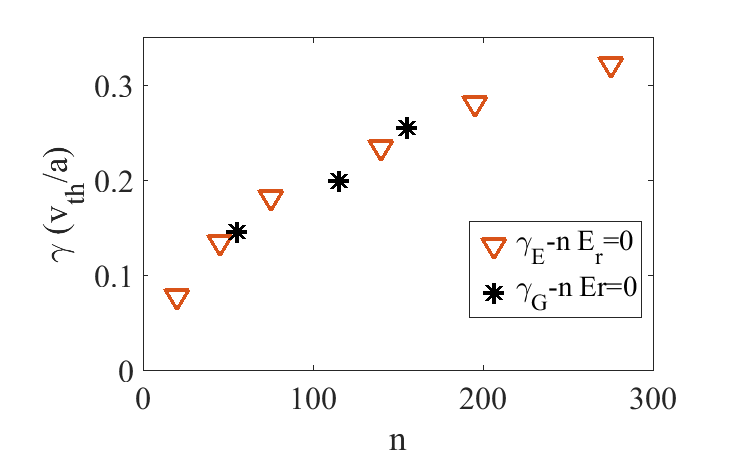}
	\caption{Linear growth rate in a set of linear simulations with EUTERPE and GENE-3D in the W7-X configuration using the profiles from Figure \ref{fig:profilesMcMillan}.}
	\label{figGRW7XMcMillan}
\end{figure}
%******************************************* 
As for the simulations using the profiles from Figure \ref{fig:profiles}, we run a set of simulations with increasing resolutions with both codes. We used $n_{\theta}=64,128,512,768,1024$ for EUTERPE and $n_y=64, 128,256$ for GENE-3D. The largest resolution simulation with EUTERPE ($n_{\theta}=1024$) resolves modes up to $m \sim 500$ ($n \sim 590$), well above the most unstable mode found, around $n \sim 280$. With GENE-3D, modes up to $n \sim 640$ can be resolved with the largest resolution simulation. The linear growth rates of the most unstable modes found in these simulations are shown in Figure \ref{figGRW7XMcMillan}.

With the profiles from Figure \ref{fig:profilesMcMillan} we  run nonlinear simulations with adiabatic electrons using both codes.  The simulation domain is restricted to $0.1 < r/a<0.9$ and we used exactly the same value for the heating source  in both codes, which was set close to one tenth of the maximum growth rate in the linear simulations, $\gamma_m \sim 0.3 ~v_{th}/a$. 
Figure \ref{fig:2DPotinNLW7XEr0} shows the density at two toroidal angle positions with $\phi=0$ and $\phi= \pi/5$ in two nonlinear   simulations carried out in W7-X using these profiles. % from Figure \ref{fig:profilesMcMillan}. 
In nonlinear simulations, the electrostatic potential is dominated by a large-amplitude zonal component. In Figure~\ref{fig:2DPotinNLW7XEr0}, we show the density instead of the electrostatic potential to better appreciate the small scale turbulent structures. The similarity between the EUTERPE and GENE-3D results is clear. 
%*******************************************
\begin{figure}[!ht]
	\centering
	\includegraphics[width=6.25cm,  trim=260 10 50 10, clip, angle=0]{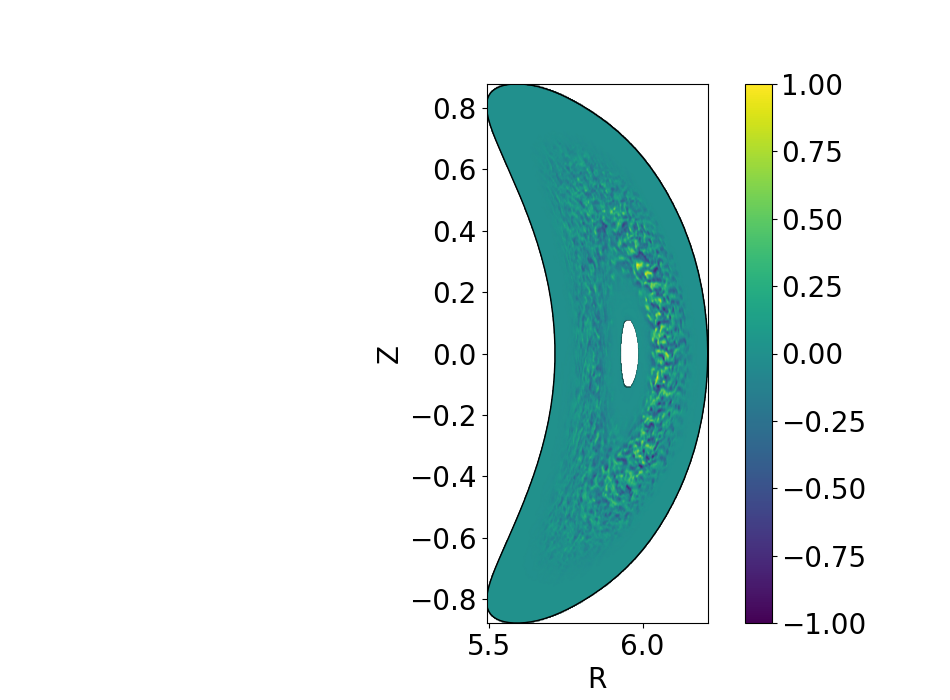}
	\includegraphics[width=8.75cm,  trim=0 -50 50 0, clip, angle=0]{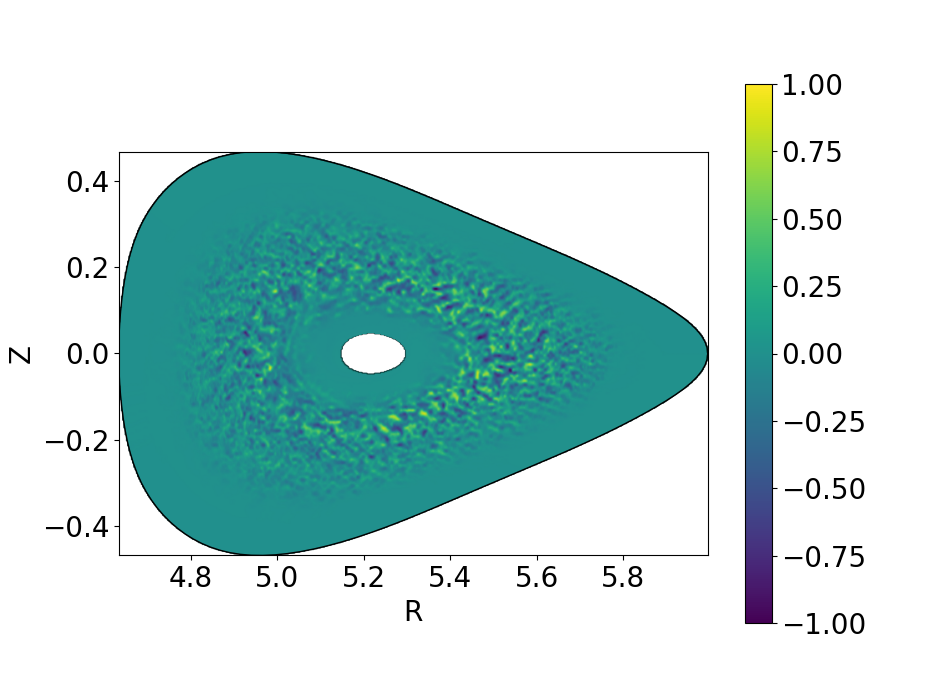}
	\\
	\hspace{-.85cm}
	\includegraphics[width=6.25cm,  trim=260 10 50 10, clip, angle=0]{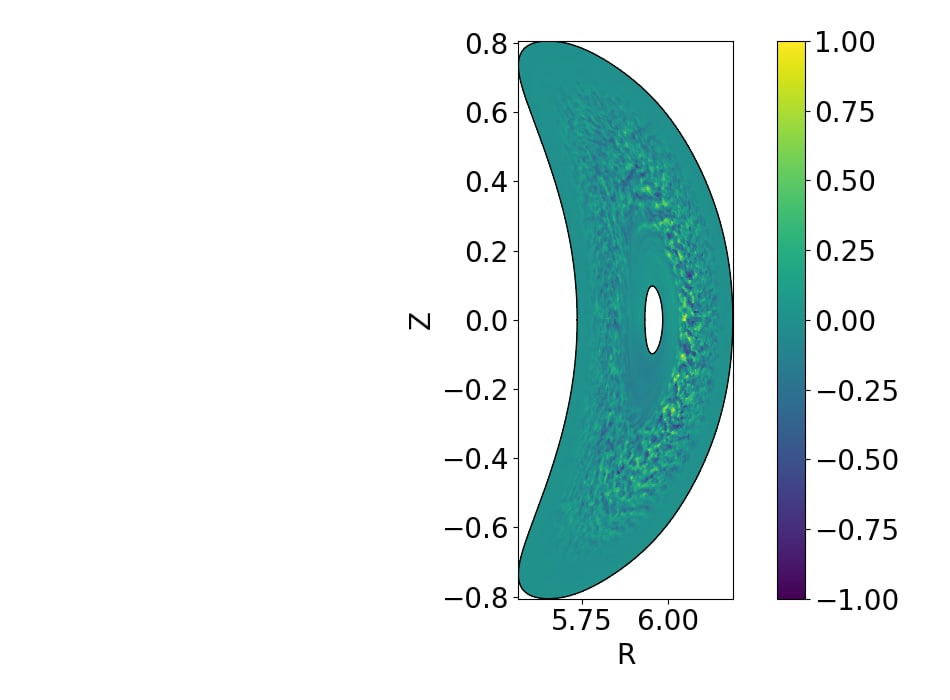}
	\includegraphics[width=8.75cm,  trim=0 -50 50 0, clip, angle=0]{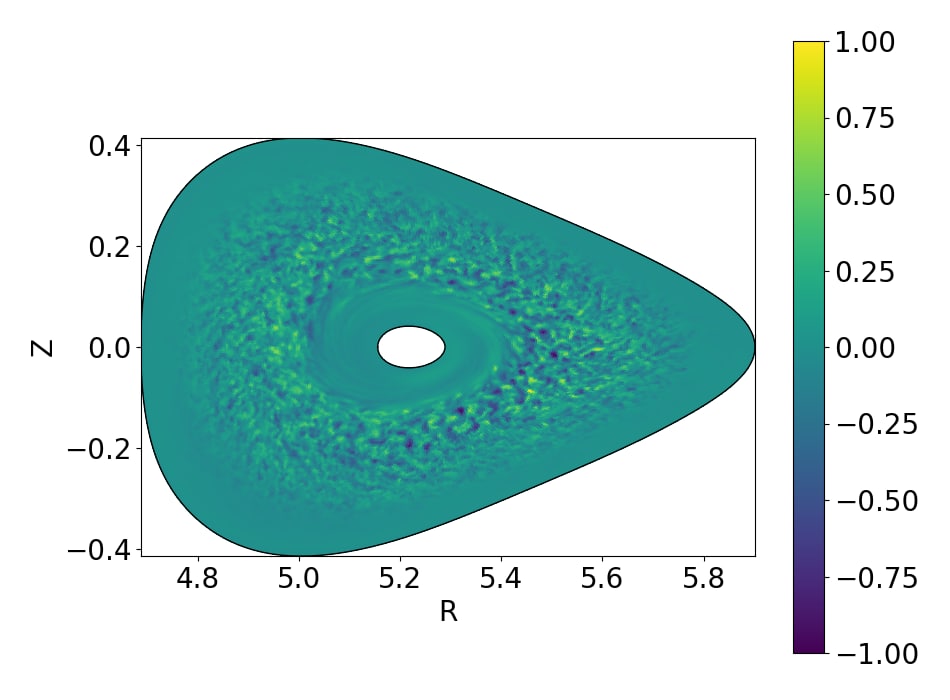}
	\\
	\caption{Density (normalized to maximum) at toroidal angles $\phi=0$ (left) and $\phi=\pi/5$ (right) for two nonlinear simulations carried out with EUTERPE (top) and GENE-3D (bottom) in the W7-X configuration using the profiles from Figure \ref{fig:profilesMcMillan}.}
	\label{fig:2DPotinNLW7XEr0}
\end{figure}
 For these simulations we calculated the surface-integrated ion heat flux, which is shown in Figure \ref{fig:heatFluxesW7X} vs radial position. 
 The shadings in Figure  \ref{fig:heatFluxesW7X} represent an uncertainty margin, which is estimated  in each case as the standard deviation of the instantaneous heat flux time trace at each radial position, computed after nonlinear saturation in a time window in which a quasi-steady state is reached.
		%*******************************************
		\begin{figure}[!h]
			\centering
			\includegraphics[width=7.5cm,  trim=30 10 50 0, clip, angle=0]{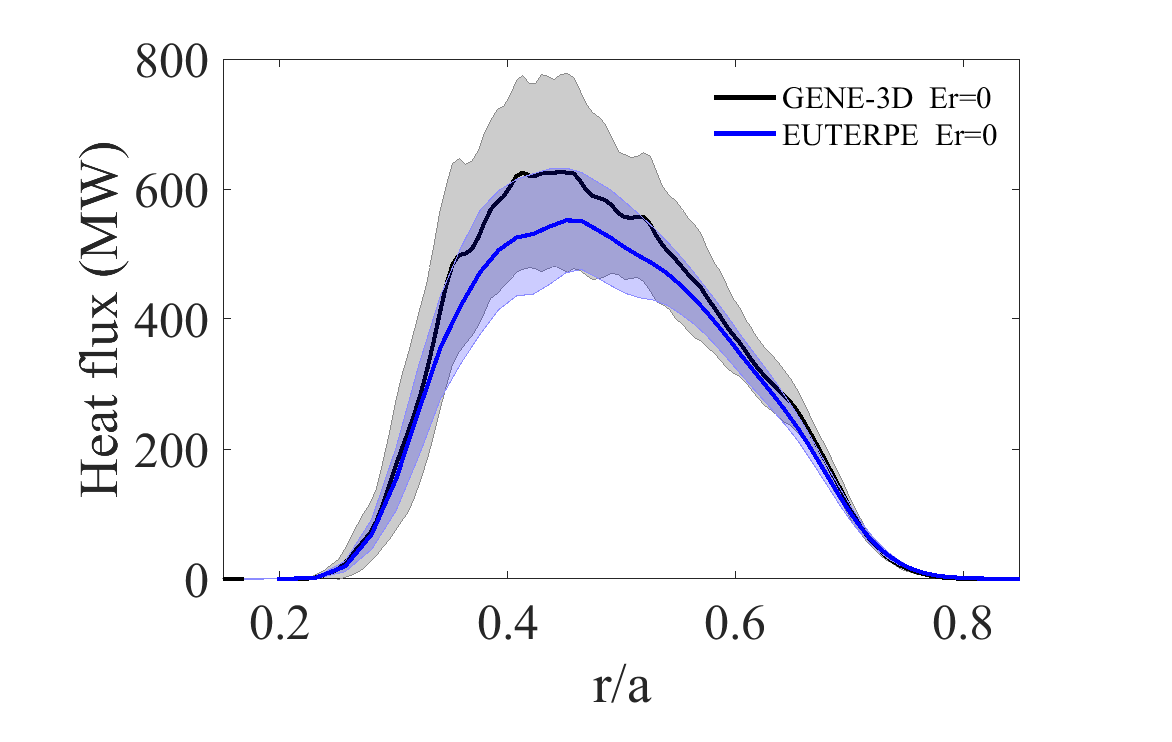}
			\caption{Surface-integrated turbulent ion heat flux vs radial position in nonlinear simulations in W7-X using the profiles from Figure \ref{fig:profilesMcMillan}.}
			\label{fig:heatFluxesW7X}
		\end{figure}
		%*******************************************
It is clear from the figure that the agreement between both codes is very good, the heat flux given by each code coinciding with that of the other one within the error margin. Extensive convergence tests were carried out in both codes, by changing the resolutions in the radial direction and also in the poloidal and toroidal angles. Radial resolutions $n_s=128$ and $n_s=144$ were used and the results were almost exactly the same. In EUTERPE, simulations with $n_{\theta}=256,384,512,768$ were carried out. The resolution in the toroidal direction was set according to the resolution in the poloidal one as to capture near-resonant modes with $n \approx \iota m$.  The heat flux showed to be well converged already for $n_{\theta}=512$. This is explained because, unlike the growth rates that peak around $n=300, k_{\theta}\rho_i\approx 3$, the heat fluxes peak at $k_{\theta}\rho_i <  1$.  Tests of convergence with the number of markers were also studied in EUTERPE, showing convergence for $N_{m}/N_{gc} \sim 20$, with $N_{gc}$ the number of grid cells, which corresponds to a ratio $N_{m}/N_{modes} \sim 350$. Note that the ratio of markers to modes is more relevant that the ratio of markers per grid cell as a measure of the quality of the simulations \cite{Bottino2007}.

These nonlinear simulations required around 24 hours walltime using 64 nodes in Marconi, which means around 70 kh core hours, to get a saturated state sufficiently long as to extract values of physical quantities. The computational cost with both codes is similar. 

%*******************************************
%*******************************************
%*******************************************
\section{Localization of instabilities and turbulence over the flux surface}\label{secLocalization}

%*******************************************
%*******************************************
Having shown the good agreement between EUTERPE and GENE-3D codes in both linear and nonlinear simulations, in  LHD and W7-X configurations, we now turn our view to some specific aspects of the plasma turbulence in stellarators for which the global codes are particularly suited.

A strong  localization of linear instabilities has been previously reported in global  simulations in different stellarators \cite{Nadeem01,Kornilov04, Riemann16, Sanchez16, Sanchez19}. Saturated ITG turbulence has also been reported  to be strongly localized in full-surface simulations with GENE \cite{Xanthopoulos14, Xanthopoulos20}. In this section, we study the localization of instabilities and turbulence over the flux surface with EUTERPE and GENE-3D and compare the results of both codes in this respect. Differences with results in full-surface simulations previously reported will be shown. 

%*******************************************
\subsection{Localization of instabilities}\label{secLocalizationInstab}
%
%*******************************************
We start looking at the localization of instabilities in the linear simulations for LHD studied in sections \ref{secCompLinearSimsLHD} to \ref{secCompLinearSimsW7X}. Figure \ref{fig:LocalPotLHD} shows the amplitude of the electrostatic potential (normalized to its maximum value on the flux surface) at the flux surface at $r/a=0.5$ vs toroidal and poloidal angles for the simulations from sections \ref{secCompLinearSimsLHD} to \ref{secCompLinearSimsW7X} in LHD carried out with EUTERPE and GENE-3D.		
		%*******************************************
		\begin{figure}[!h]
			\centering
			
			\includegraphics[width=7.5cm,  trim=30 10 50 24, clip, angle=0]{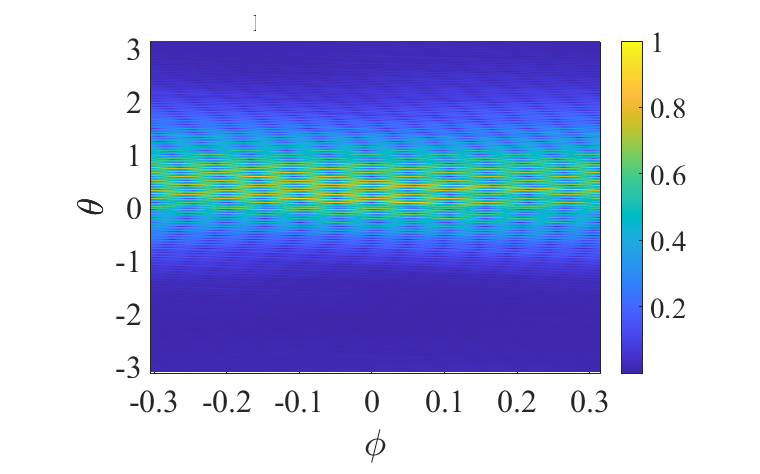}
			\includegraphics[width=7.5cm,  trim=30 10 50 24, clip, angle=0]{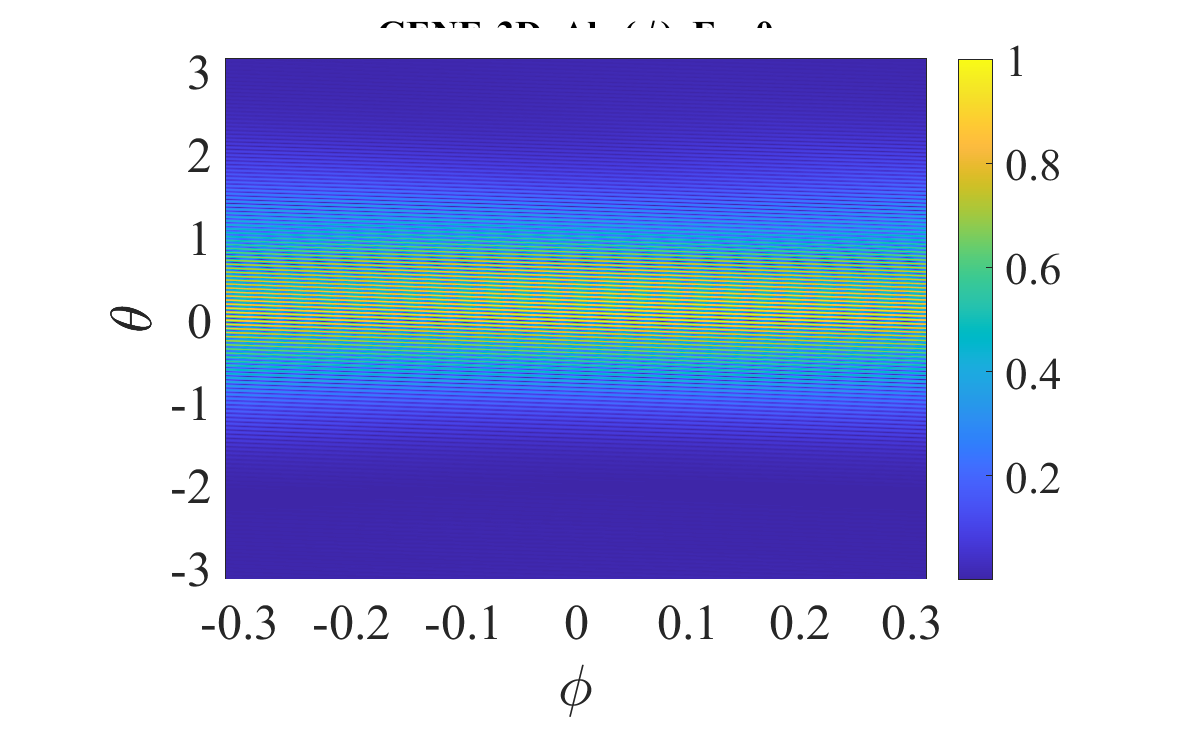}\\
			\includegraphics[width=7.5cm,  trim=30 10 50 24, clip, angle=0]{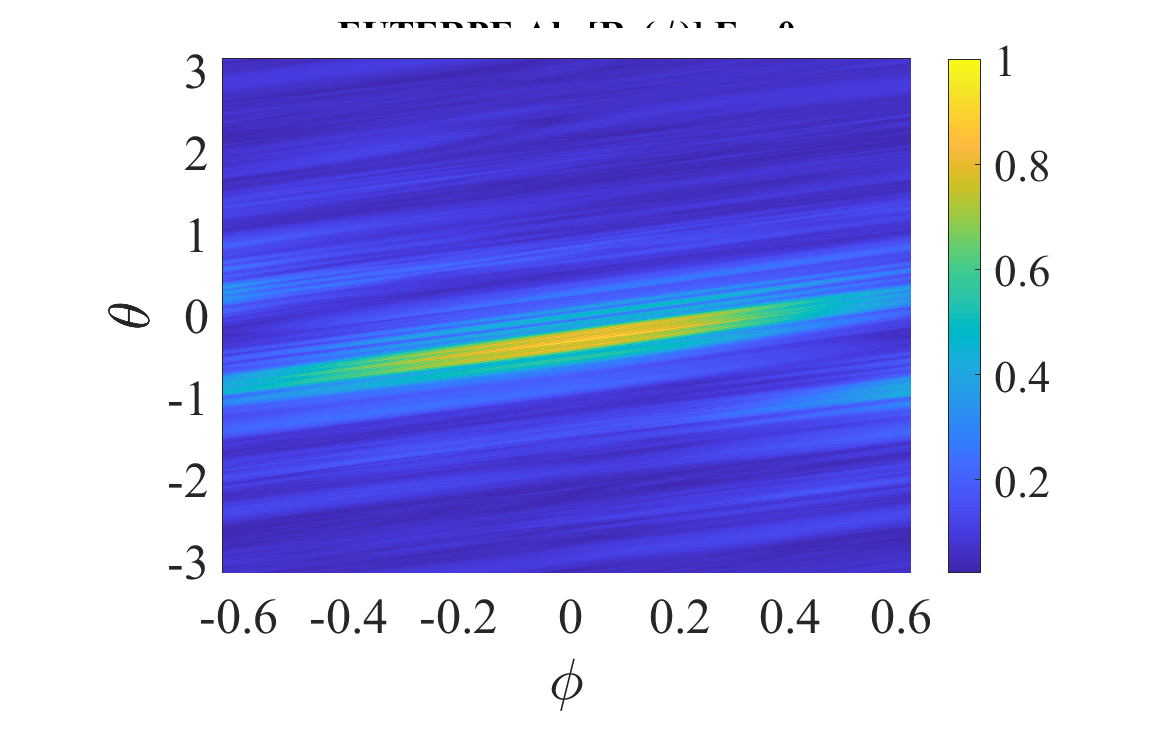}
			\includegraphics[width=7.5cm,  trim=30 10 50 24, clip, angle=0]{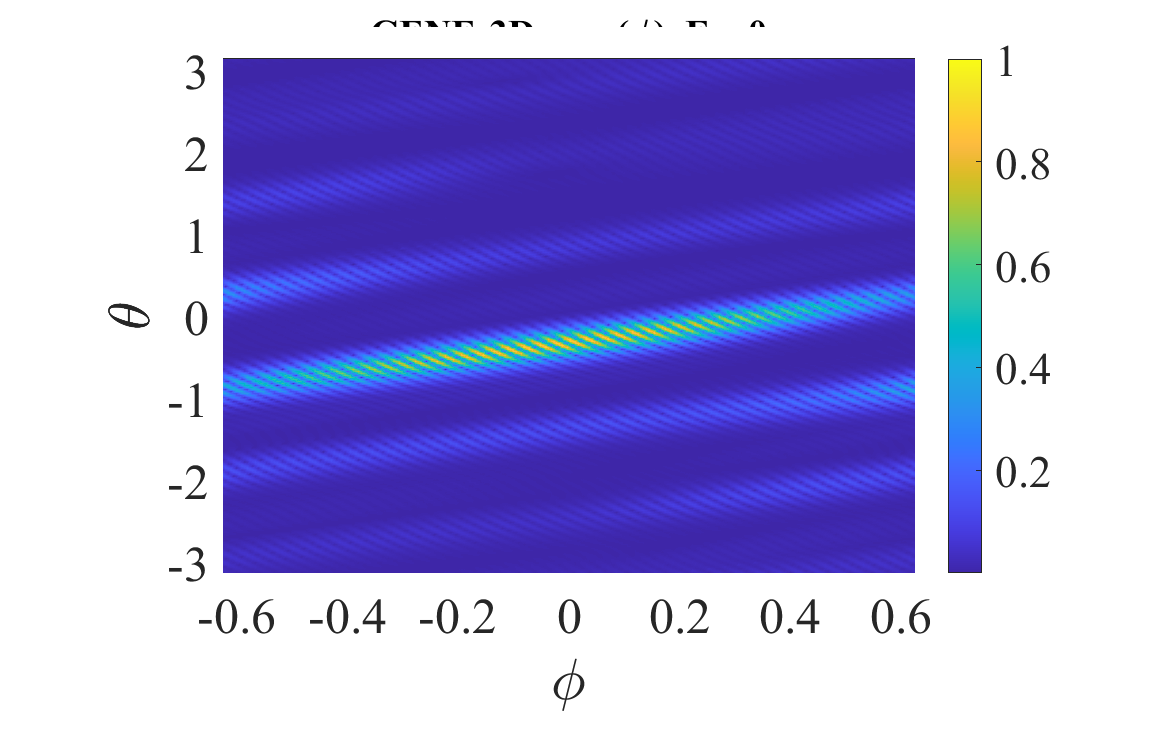}
			\caption{Amplitude of the electrostatic potential (normalized to its maximum value) at the flux surface at $r/a=0.5$ versus toroidal ($\phi$) and poloidal ($\theta$) angles for the linear simulations in the LHD configuration presented in Section \ref{secCompLinearSimsLHD} (top) and for the simulations in W7-X configuration from Section \ref{subsecW7XLinAE} (bottom). The left panels correspond to the simulations with EUTERPE while those on the right show results with GENE-3D.}
			\label{fig:LocalPotLHD}
		\end{figure}
		%*******************************************
A clear localization of the maximum amplitude of the instability is found by both codes. The spatial region with  maximum amplitude of the potential perturbation is aligned with the field lines. Similar results are obtained if we plot the density instead of the electrostatic potential. Note that the simulations are carried out with adiabatic electrons and then the density is closely related with the potential.  
The potential is shown for a fixed time instant. 
Note that in this kind of simulations, the amplitude of the perturbation grows exponentially in time which makes that one or a few modes dominate after a time of simulation.
The resolution in the EUTERPE simulation was $n_{\theta}=256$ and for GENE3-D $n_y=64$. 
Figure \ref{fig:LocalPotLHD} shows also the electrostatic potential for the linear simulations in the W7-X configuration presented in Section \ref{subsecW7XLinAE}. In this case, the simulation with EUTERPE has resolution $n_{\theta}=768$ and that with GENE-3D with $n_y=128$. The density shows similar localization.
Comparing the simulations for LHD and W7-X shown in this figure we see  that the region of maximum instability is aligned with the field line, the spatial localization is more pronounced (smaller poloidal extent of the region of large perturbation amplitude) in W7-X than in LHD, and the agreement between the codes in respect to the localization of instabilities is very good in both cases.

%	%*******************************************
	%*******************************************
	\begin{figure}[!h]
		\centering
		\includegraphics[width=7.5cm,  trim=30 10 50 20, clip, angle=0]{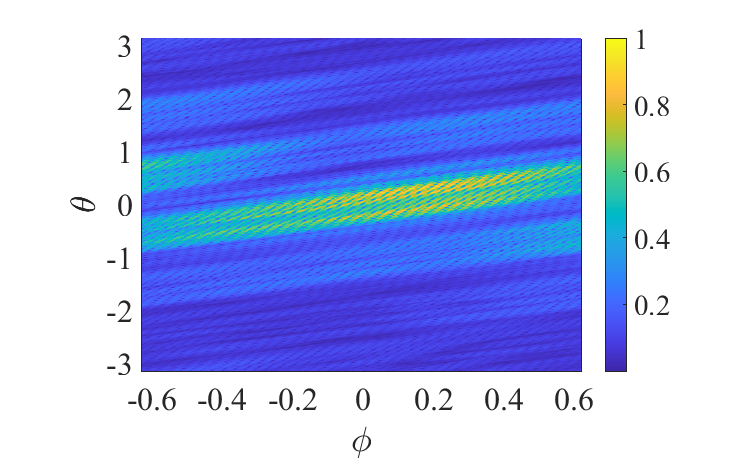}
		\includegraphics[width=7.5cm,  trim=30 10 50 20, clip, angle=0]{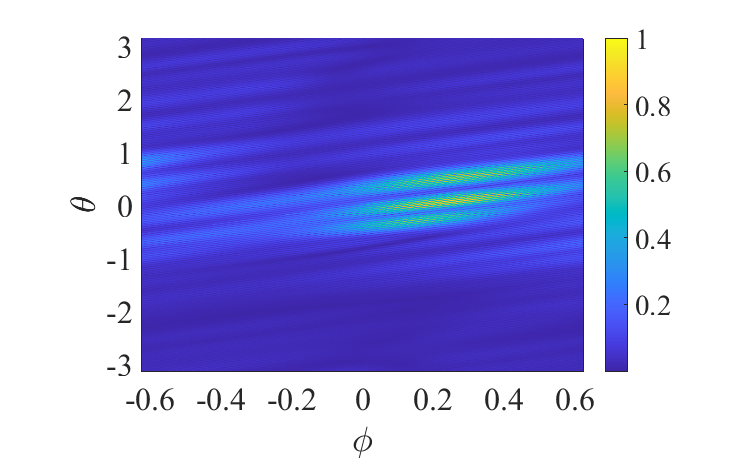}
			\includegraphics[width=7.5cm,  trim=30 10 50 20, clip, angle=0]{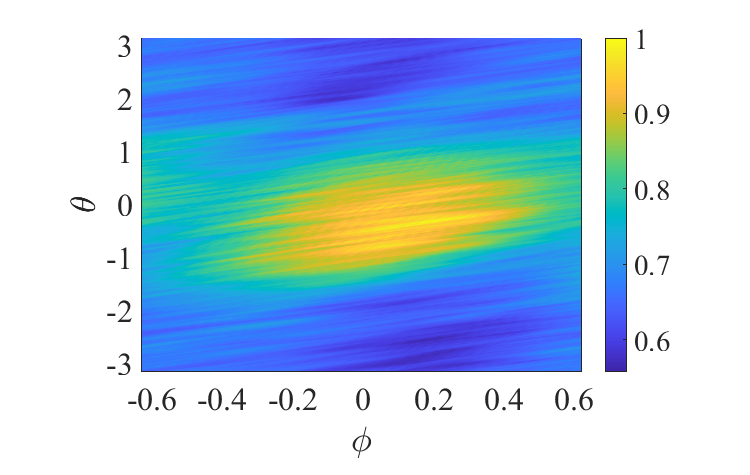}
		\includegraphics[width=7.5cm,  trim=30 10 50 20, clip, angle=0]{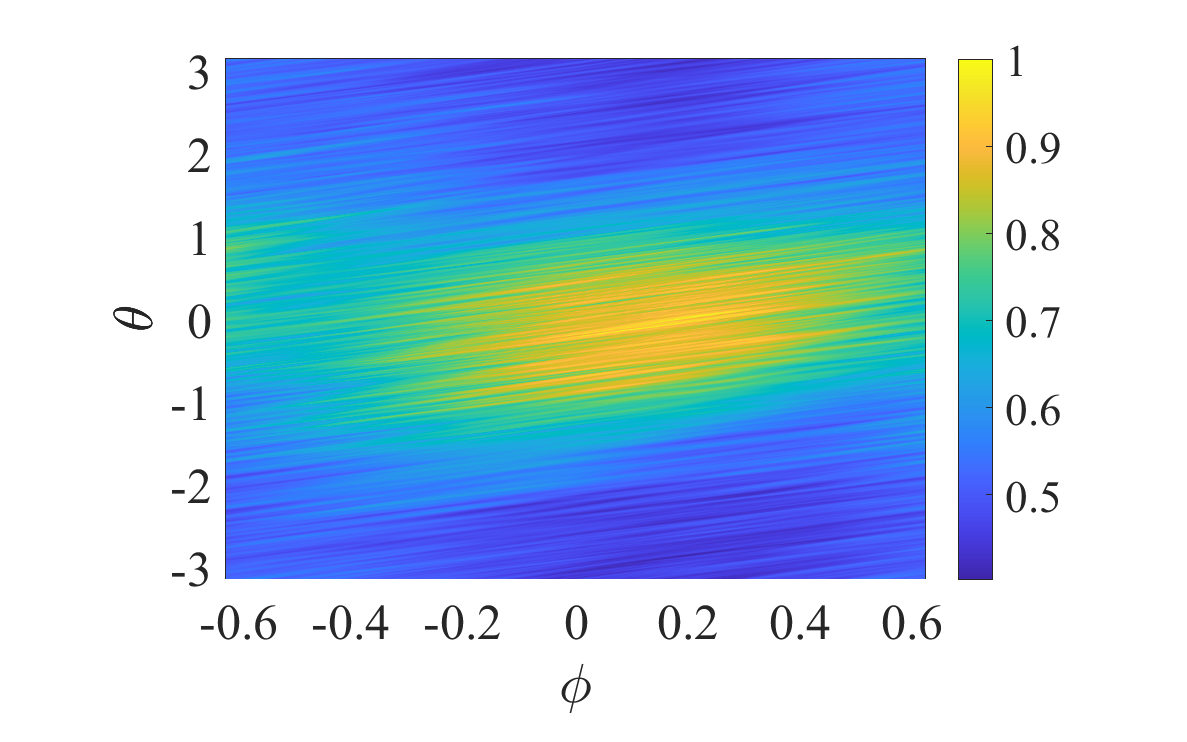}
		\caption{Density (normalized to its maximum value) at the flux surface at $r/a=0.5$ versus toroidal and poloidal angles for the linear (top) and nonlinear (bottom) simulations for the W7-X configuration with EUTERPE (left) and GENE-3D (right)  presented in Section \ref{secNonLinW7X}. The top panels correspond to linear simulations and the bottom ones to nonlinear simulations. Density at a time instant is shown in the top panels while the \textit{rms} of the density, extracted in a long time interval during the saturated phase, is shown for the nonlinear (bottom) panels. The EUTERPE simulations where carried out with $n_{\theta}=512$ and the GENE-3D ones with $n_y=128$.}
		\label{fig:LocalPotW7X2}
	\end{figure}
	%*******************************************
%*******************************************
\subsection{Localization of turbulence}\label{secLocalizationTurb}

%*******************************************
Now we use the nonlinear simulations presented in Section \ref{secNonLinW7X} to study to what extent the localization of linearly unstable modes remains after the nonlinear saturation of turbulence. In nonlinear simulations we have a long quasi-steady state in which we can extract time-averaged measures for this localization. We will look at the density instead of the potential as for the later strong zonal components appear with a long time evolution which dominates the spectrum and then the localization is less clearly appreciated if the zonal component is not removed first. Density does not show that strong zonal component and the localization is more readily appreciated.

Figure \ref{fig:LocalPotW7X2}-top shows the density (normalized to its maximum value over the flux surface)  at a fixed time instant at the end of the linear simulations described in Section \ref{secNonLinW7X}. In addition, the root mean square (\textit{rms}) value of of the density extracted in a long time during the quasi-steady-state phase of saturated turbulence is shown in the bottom panels of the same figure, which is more appropriate for comparison than instant values. In this case, we show data corresponding to the EUTERPE simulation carried out with resolution $n_{\theta}=512$, which is the resolution for which we find numerical convergence in resolution, and for the simulations with $n_y=128$ for GENE-3D. Note that the temperature profiles used in these simulations (see Figure \ref{fig:profilesMcMillan}) allow more modes at different radial locations are excited as compared to the profiles used in sections \ref{secCompLinearSimsLHD} and \ref{secCompLinearSimsW7X} (shown in Figure \ref{fig:profiles}) and the localization of instabilities in Figure \ref{fig:LocalPotW7X2}-top is not as strong as that in Figure \ref{fig:LocalPotLHD}-bottom.   

It is clear that the localization of turbulent fluctuations is not as large as that of the linear mode amplitude. The spatial region of the flux surface covered by the largest amplitude of fluctuations is significantly wider in the direction perpendicular to the field line in the nonlinear cases as compared to the linear ones. {The weaker localization of nonlinear fluctuations can be related to the different wavenumber spectrum that emerges after nonlinear saturation, whose spectral power is distributed across a wider number of wavenumbers as compared to the linear spectrum of unstable modes, dominated by a few modes. It can also be related to the influence of global elements, such as the zonal flow, which affects the nonlinear saturation, connecting perturbations localized over the flux surface, while the linear instabilities are more affected by the local magnetic geometry.}
A good agreement between the codes is also found in this respect. The agreement in the location of maximum fluctuations can be addressed quantitatively if we attend at the numerical value of the ratio of maximum to minimum \textit{rms} over the flux surface. Both codes give a value around 2.5 for this ratio, which is significantly smaller than that reported in \cite{Xanthopoulos14}, around 9. The poloidal localization, as measured by the inverse width of the maximum-amplitude spot in the poloidal direction, obtained in the present work is also significantly smaller than in reference \cite{Xanthopoulos14}.

Summarizing, a strong localization of the ITG instabilities has been found with both GENE-3D and EUTERPE codes in LHD and W7-X. These results are in good qualitative agreement with early results reported in \cite{Nadeem01} and those reported in W7-X \cite{Kornilov04,Riemann16} and TJ-II \cite{Sanchez19}. The localization is much more pronounced in W7-X than in LHD, as shown in Figure \ref{fig:LocalPotLHD}}. 
However, the localization of turbulent fluctuations is less pronounced than that of linear instabilities, as shown in Figure \ref{fig:LocalPotW7X2}, and significantly smaller than previously reported in full-surface simulations with GENE \cite{Xanthopoulos14}.
%
%*******************************************
%*******************************************
%
\section{Influence of a background radial electric field}\label{secElecField}
%
%*******************************************
%*******************************************
%
This section is devoted to analyze the influence of a background long-wavelength radial electric field in both linear instabilities and saturated turbulence. This is another aspect particularly suited for global codes as an electric field does not have other effect than producing a Doppler shift in local flux tube simulations and full-surface or global simulations are required in order to properly account for its effect. 

In addition to the simulations presented in sections \ref{secCompLinearSims} and \ref{secNonLin}, we run a set of simulations with each code including a background radial electric field. The electric field is constant in radius and several values of the field strength were considered.
%*******************************************

\subsection{Influence of $E_r$ on linear growth rates}\label{subSecElecFieldLinGR}

%*******************************************
We will start studying the stabilization of linear ITG modes in LHD. We run a set of linear simulations using the same density and temperature profiles shown in Section \ref{secCompLinearSimsLHD} and including a constant radial electric field with different strengths $E_r=6~\rm{kV/m},-6~\rm{kV/m}, 12~\rm{kV/m},-12~\rm{kV/m}$.

For each simulation, the most unstable mode was identified, their toroidal and poloidal mode numbers were extracted, and the time evolution of electrostatic potential for this mode was fit to an  exponentially growing  oscillation model, and the growth rate and frequency were extracted from the fit. The results of this analysis, together with those of the simulations without electric field presented in Section \ref{secCompLinearSimsLHD} are compiled in table \ref{tab:NMGRFLHD}.

\begin{table}[!h]
	\centering
	\caption{Summary of results (most unstable mode, growth rate and frequency) in EUTERPE and GENE-3D linear simulations in LHD. In these simulations the reference temperature is $T=4.25~\rm{eV}$, $a=0598~\rm{m}$, $k_{\theta}=m/r$ and $\rho_i=\sqrt{2 m T}/eB$. 	$\gamma$ and $\omega$ are given in units of $v_{thi}/a$. Positive/negative frequency means propagation in the ion/electron diamagnetic drift direction. }
	\label{tab:NMGRFLHD}	
	\begin{tabular}{|c|c|c|c|}
		\hline
		$\mathbf{E_r}$&			&		\multirow{2}{*}{\bf{EUTERPE}}		&	\multirow{2}{*}{\bf{GENE-3D}}	\\
		$(\rm{kV/m})$&			&			&	\\
		\hline
		\multirow{4}{*}{$0$}&	n, m			&		30, 64		&	30, 65 	\\
		%	\hline
		\cline{2-4}
		&	{$k_{\theta}\rho_i$}			&	{	0.57}		&	{0.58 	}\\
		%	\hline
		\cline{2-4}
		&		$\gamma $	&	$	0.21 $		&			$	0.20 $	\\
		%	\hline
		\cline{2-4}
		&	$\omega	$		&	$0.39 $ 		&		$0.40 $	\\
		\hline
		\hline
		\multirow{4}{*}{$+6$}&	n, m		&	30,	65	&	30, 65		\\
		%	\hline
		\cline{2-4}
		&	{$k_{\theta}\rho_i$}			&	{	0.58}		&	{0.58 	}\\
		%	\hline
		\cline{2-4}
		&	$\gamma$		&	$0.19$		&	$0.18 $		\\
		%	\hline
		\cline{2-4}
		&		$\omega	$	&	$0.92$		&	$0.93$		\\
		\hline
		\hline
		\multirow{4}{*}{$-6$}&		n, m	&	30, 64		&			30, 65		\\
		%	\hline
		\cline{2-4}
		&	{$k_{\theta}\rho_i$}			&	{	0.57}		&	{0.58 	}\\
		%	\hline
		\cline{2-4}
		&	$\gamma	$		&	$0.21$			&	$0.2$			\\
		%	\hline
		\cline{2-4}
		&	$\omega	$		&	$0.15$			&		$0.14$		\\
		\hline
		\hline
		\multirow{4}{*}{$+12$}&	n, m		&	30, 66		&30,	65		\\
		%	\hline
		\cline{2-4}
		&	{$k_{\theta}\rho_i$}			&	{	0.59}		&	{0.58	}\\
		%	\hline
		\cline{2-4}
		&	$\gamma$		&		$0.18$		&		$0.17$		\\
		%	\hline
		\cline{2-4}
		&		$\omega	$	&	$1.49$		&		$1.49$	\\
		\hline
		\hline
		\multirow{4}{*}{$-12$}&	n, m		&		30, 64	&	30, 64		\\
		%	\hline
		\cline{2-4}
		&	{$k_{\theta}\rho_i$}			&	{	0.57}		&	{0.57 	}\\
		%	\hline
		\cline{2-4}
		&		$\gamma	$	&		$0.22$		&		$0.20$	\\
		%	\hline
		\cline{2-4}
		&		$\omega$	&		$-0.67$		&		$-0.67$			\\
		\hline
	\end{tabular}
	
\end{table}

The agreement between codes on the toroidal mode number, $n$, is perfect, while for the poloidal mode number, $m$, only a difference in $\pm 1$ appears in some cases.

For positive values of the electric field a reduction of the growth rate of the most unstable mode is found (linear stabilization by $E_r$). The reduction of $\gamma$ increases with the field strength ($ 10 \%$ for $E_r=6~kV/m$ and $14\%$ for $E_r=12~kV/m$). The frequency of the mode increases with the field strength ({Doppler shift). 
	For negative electric fields, the growth rate is only slightly modified (increased) in this case by ~0.2\%. With respect to the frequency of the mode, it is reduced for $E_r=-6~kV/m$ and then it increases for $E_r=-12~kV/m$. This can be explained by the fact that the negative electric field induces a rotation in the direction opposite to the ion diamagnetic drift; then, for $E_r=-6~kV/m$ the velocity of propagation of modes,  in the laboratory frame of reference, is reduced with respect to the case without electric field, and for $E_r=-12~kV/m$ the direction of rotation of modes changes sign.
	
	Following the set of linear simulations without electric field presented in Section \ref{secNonLinW7X}, in W7-X we run a set of simulations with EUTERPE and GENE-3D including $E_r$ and using increasing resolutions in the angles. For each simulation, we selected the most unstable mode and extracted the growth rate. The values of the electric field strength considered in this case are $E_r=20~\rm{kV/m},-20~\rm{kV/m},40~\rm{kV/m}$. The results of growth rate of the most unstable mode are shown versus mode number in Figure \ref{fig:GRErW7X}. Different colors and symbols are used for different values of $E_r$ and code. Data for the simulations with EUTERPE are shown with subscript "E" and "G" is used for GENE-3D.

	%*******************************************
	\begin{figure}[!h]
		\centering	
		
		\includegraphics[width=8.5cm,  trim=30 10 0 0, clip, angle=0]{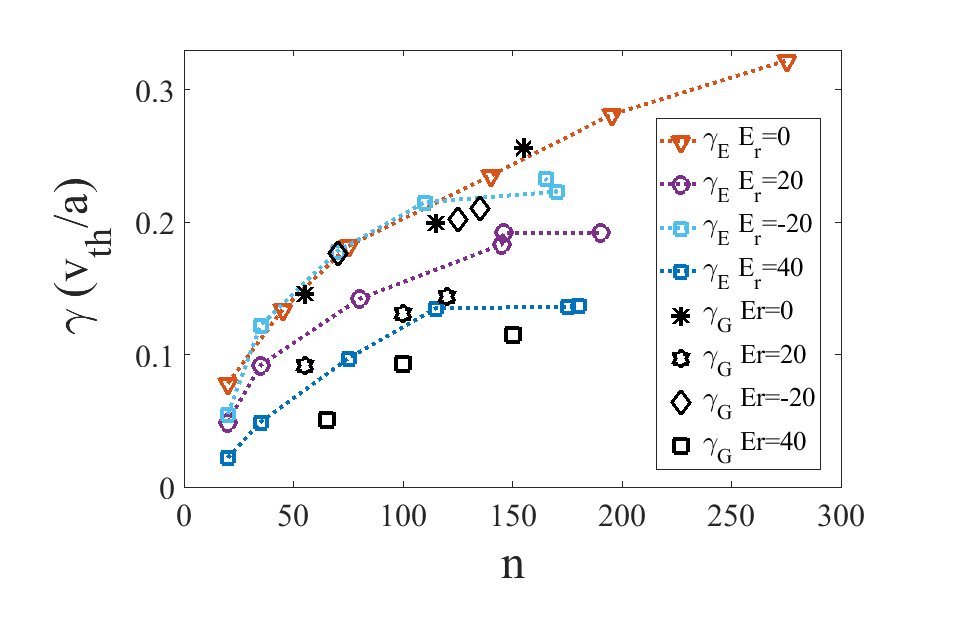}
		\caption{Growth rate of the most unstable mode in a set of simulations carried out with increasing resolutions with EUTERPE (subscript "E") and GENE-3D (subscript "G") in the W7-X configuration, and including different values of electric field. Different colors and symbols are used for different values of $E_r$ and codes (see text for a detailed description).}
		\label{fig:GRErW7X}
	\end{figure}
	%*******************************************
	
	As in the cases without electric field, there is not full agreement   between the codes on the mode number and growth rate of the most unstable mode in a point to point comparison. However, there is reasonable agreement if we compare the curves of growth rate versus mode number. The agreement is better for $E_r=0$ than for finite values of $E_r$. Positive values of the electric field $E_r=20, 40~\rm{kV/m}$ reduce the growth rate while the effect of a negative $E_r=-20~\rm{kV/m}$ is almost negligible. This is consistent with results in LHD, where the influence of a negative electric field on the growth rate was smaller than that of a positive $E_r$. The reduction of the growth rate  increases with electric field strength in both codes for positive electric fields. The reduction of growth rate observed in GENE-3D is slightly larger than that in EUTERPE simulations, however. 
	{Note that the results discussed above  contrast with those obtained in experimental plasmas \cite{EstradaNF2021}, in which a negative electric field is shown to significantly reduce the linear growth rate of ITG modes while in this work a negative electric field has a very small influence in the growth rate. The difference can be explained by the fact that those experimental plasmas show not only a significant electric field but also significant radial derivative.}

	The most unstable mode number in simulations including $E_r$ is significantly reduced as compared to the cases without electric field in both codes, which indicates a preferential stabilization of smaller-scale modes by $E_r$. This result is consistent with the observation in \cite{EstradaNF2021}, where a stabilization of the larger-wavenumber modes is observed when the experimental $E_r$ is included in simulations for the post-pellet program (180918041) in EJM magnetic configuration.

	The frequency of the modes for positive electric fields increases significantly and a reliable fit is very difficult with the available data.

%*******************************************

\subsection{Influence of $E_r$ on localization of instabilities}\label{subSecElecFieldInstab}

%*******************************************
We will now look 
at the influence of the electric field on the localization of instabilities and will concentrate on W7-X to illustrate this effect.  
The values of $E_r$ considered are such that the Mach number is comparable to that in experimental discharges in W7-X, with values of $E_r$ in the order of 10 to 20 ~{kV/m}, with a rather smaller temperature (see for instance \cite{EstradaNF2021}).

The 
density on a flux surface ar $r/a=0.5$ is shown  in Figure \ref{fig:LocalW7XwEr} vs toroidal and poloidal angles for the three linear simulations with $E_r=0, 20~\rm{kV/m},-20~\rm{kV/m}$ carried out with EUTERPE and GENE-3D.

%*******************************************
\begin{figure}[!h]
	\centering
	\rotatebox{90}{\quad\quad\quad EUTERPE}
	\includegraphics[height=3.9cm,  trim=30 10 115 23, clip, angle=0]{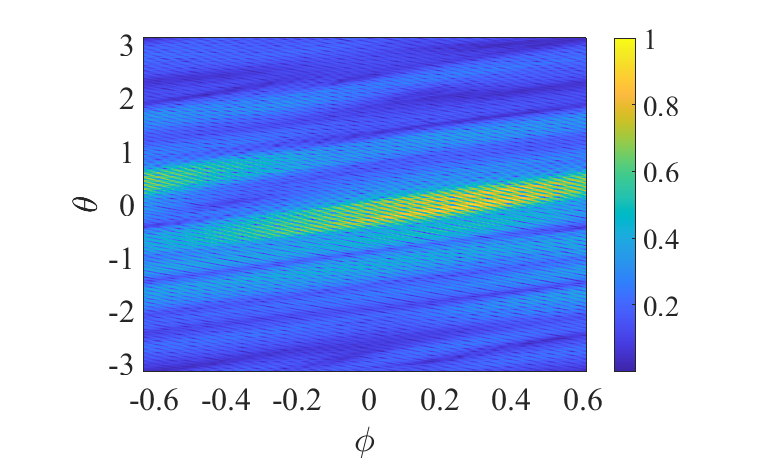}
	\includegraphics[height=3.9cm,  trim=83 10 115 23, clip, angle=0]{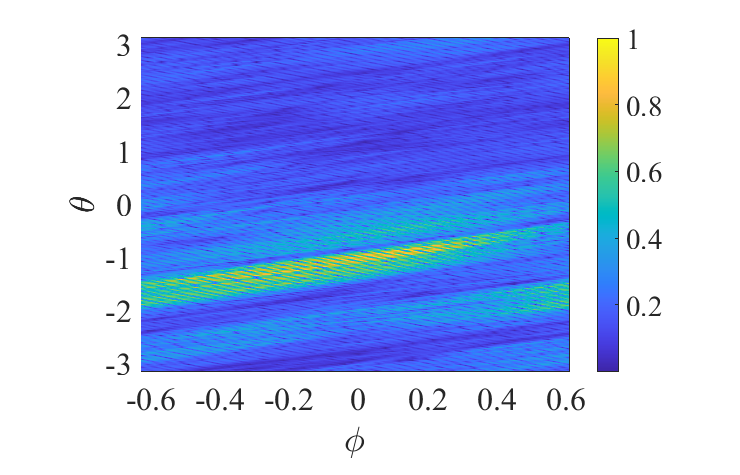}
	\includegraphics[height=3.9cm,  trim=83 10 50 23, clip, angle=0]{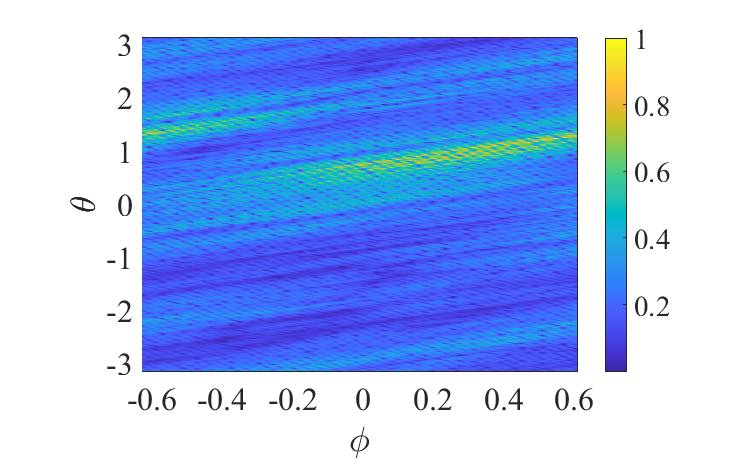}\\
	%***********************
	\rotatebox{90}{\quad\quad\quad GENE-3D}
	\includegraphics[height=3.9cm,  trim=30 10 115 23, clip, angle=0]{W7X_PhiThetaPhi_LinAE_Er0.png}
	\includegraphics[height=3.9cm,  trim=83 10 115 23, clip, angle=0]{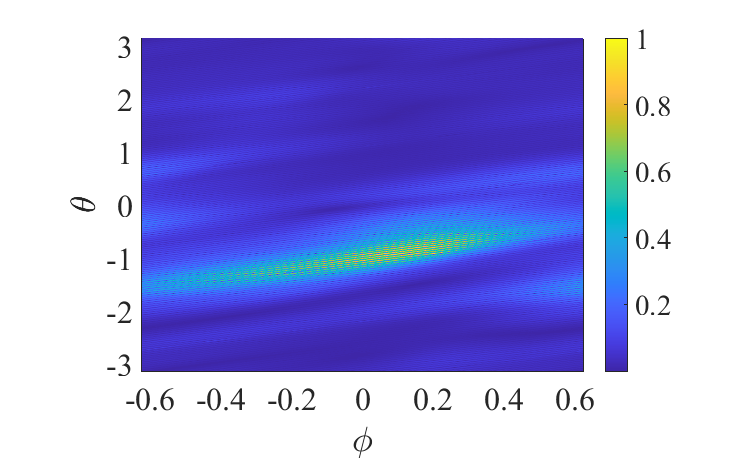}
	\includegraphics[height=3.9cm,  trim=83 10 50 23, clip, angle=0]{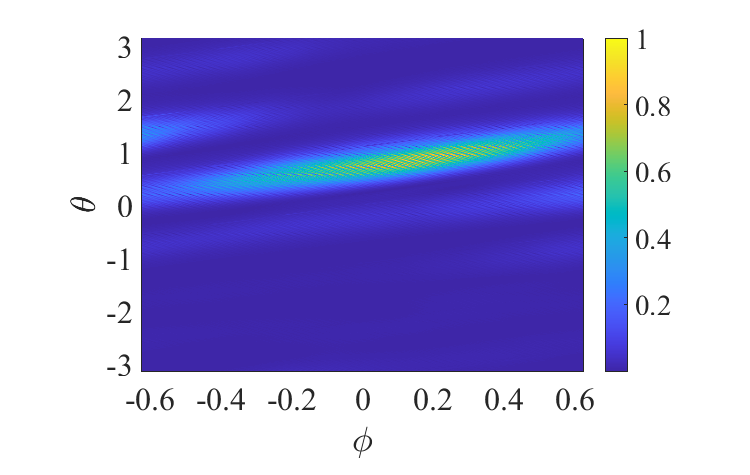}\\
	\caption{Amplitude of the density (normalized to its maximum value) at the flux surface at $r/a=0.5$ vs toroidal and poloidal angles for a set of linear simulations in the W7-X configuration with EUTERPE (top) and GENE-3D (bottom) including several values of radial electric field, $Er=0, 20~\rm{kV/m},-20~\rm{kV/m}$, from left to right. }
	\label{fig:LocalW7XwEr}
\end{figure}
%*******************************************
From the figure it is clear that the radial electric field introduces a poloidal shift of the location of maximum instability. A positive electric field produces a shift in the ion-diamagnetic drift direction (downwards in the figure) while a negative $E_r$ produces a shift in the opposite direction. A slight shift in the toroidal direction can also be appreciated that goes to the left for positive electric field and to the right for negative electric field. The poloidal shift of the location of maximum instability with $E_r$ is consistent with previous simulations with EUTERPE \cite{Riemann16}.

 \subsection{Influence of $E_r$ on localization of turbulence}\label{subSecElecFieldTurb}
 
 %*******************************************
 Now we turn to study the influence of the electric field on  the localization of maximum fluctuations in saturated turbulence. As for the linear cases, in addition to the simulations presented in Section \ref{secNonLinW7X} without external electric field  we run a set of nonlinear simulations in W7-X including a radial electric field with different signs $E_r=20~\rm{kV/m},-20~\rm{kV/m}$. 
 Figure \ref{fig:LocalW7XwErNL} shows the \textit{rms} of the density in these simulations  with EUTERPE and GENE-3D for the flux surface at $r/a=0.5$. 
 
 %*******************************************
 \begin{figure}[!h]
 	\centering
 	\rotatebox{90}{\quad\quad\quad EUTERPE}
 	\includegraphics[height=3.8cm,  trim=30 10 115 23, clip, angle=0]{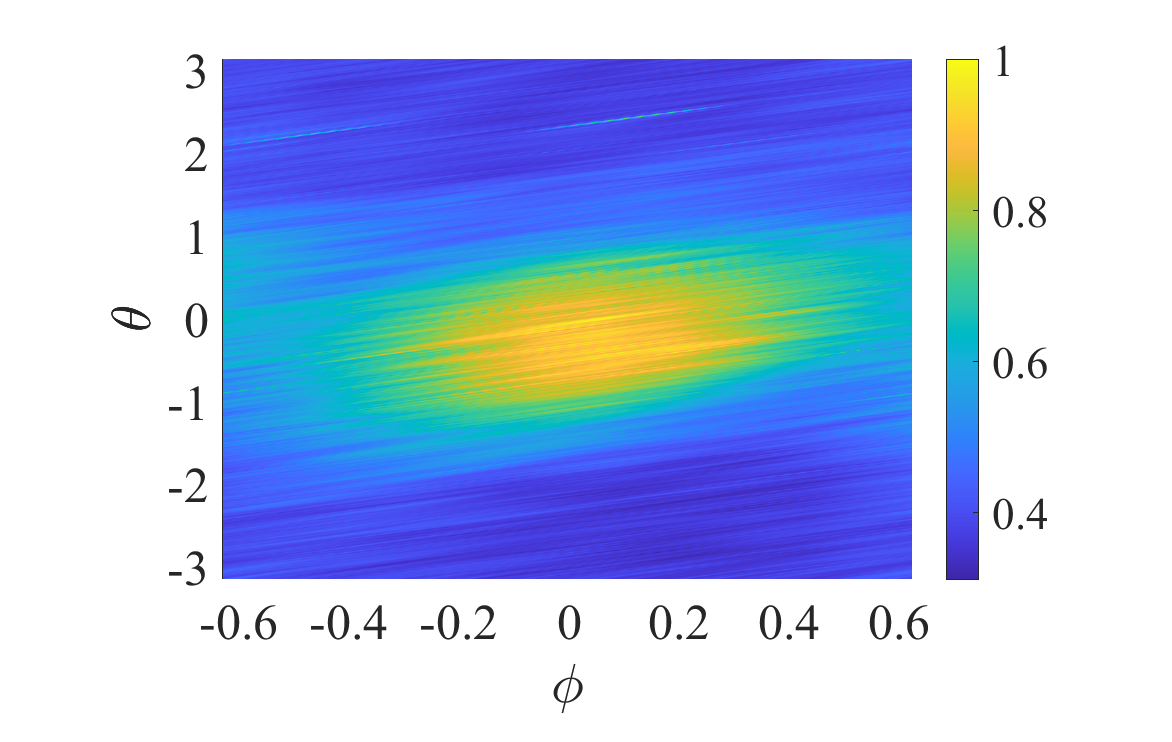}
 	\includegraphics[height=3.8cm,  trim=83 10 115 23, clip, angle=0]{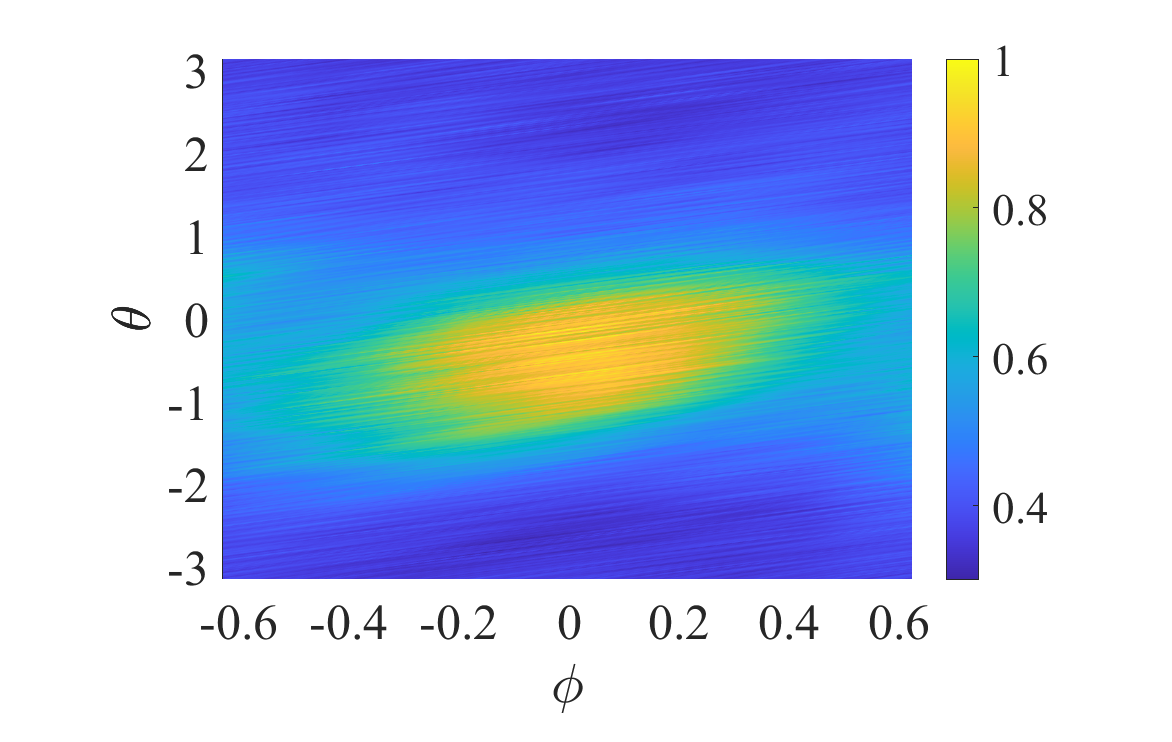}
 	\includegraphics[height=3.8cm,  trim=83 10 50 23, clip,  angle=0]{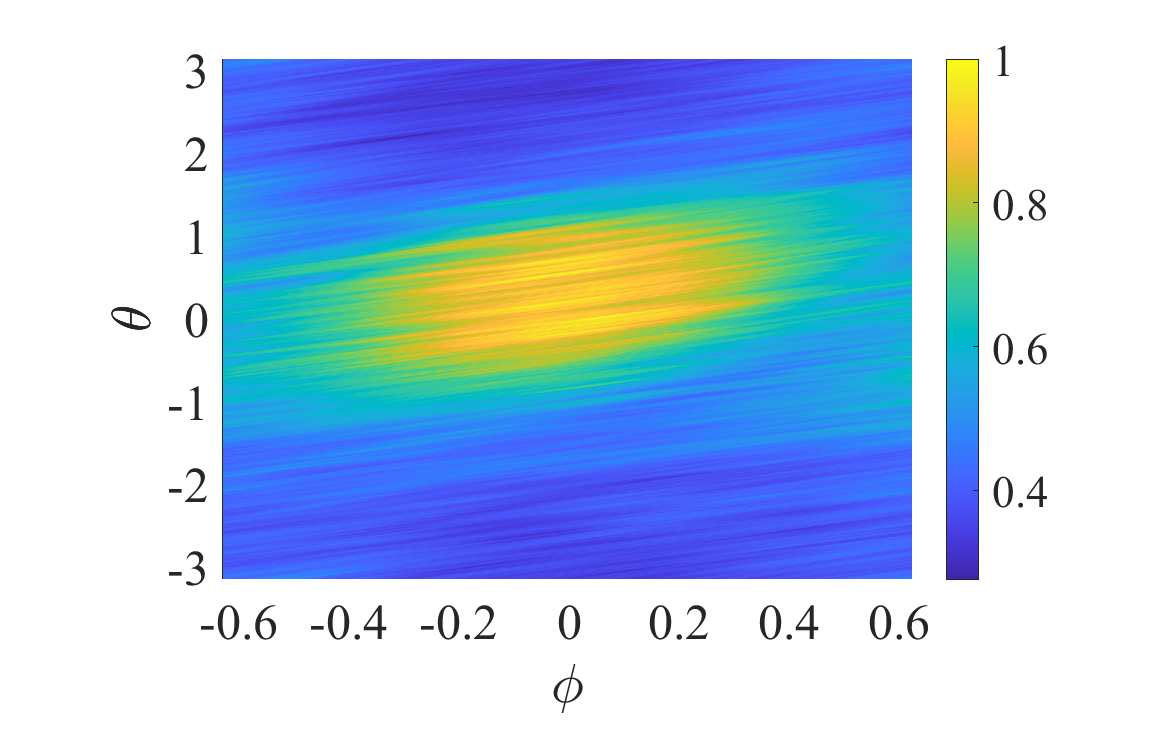}\\
 	%***********************
 	\rotatebox{90}{\quad\quad\quad GENE-3D}
 	\includegraphics[height=3.8cm,  trim=30 10 115 23, clip, angle=0]{W7X_GENE_PhiThetaDens_NLinAE_Er0HR.png}
 	\includegraphics[height=3.8cm,  trim=83 10 115 23, clip, angle=0]{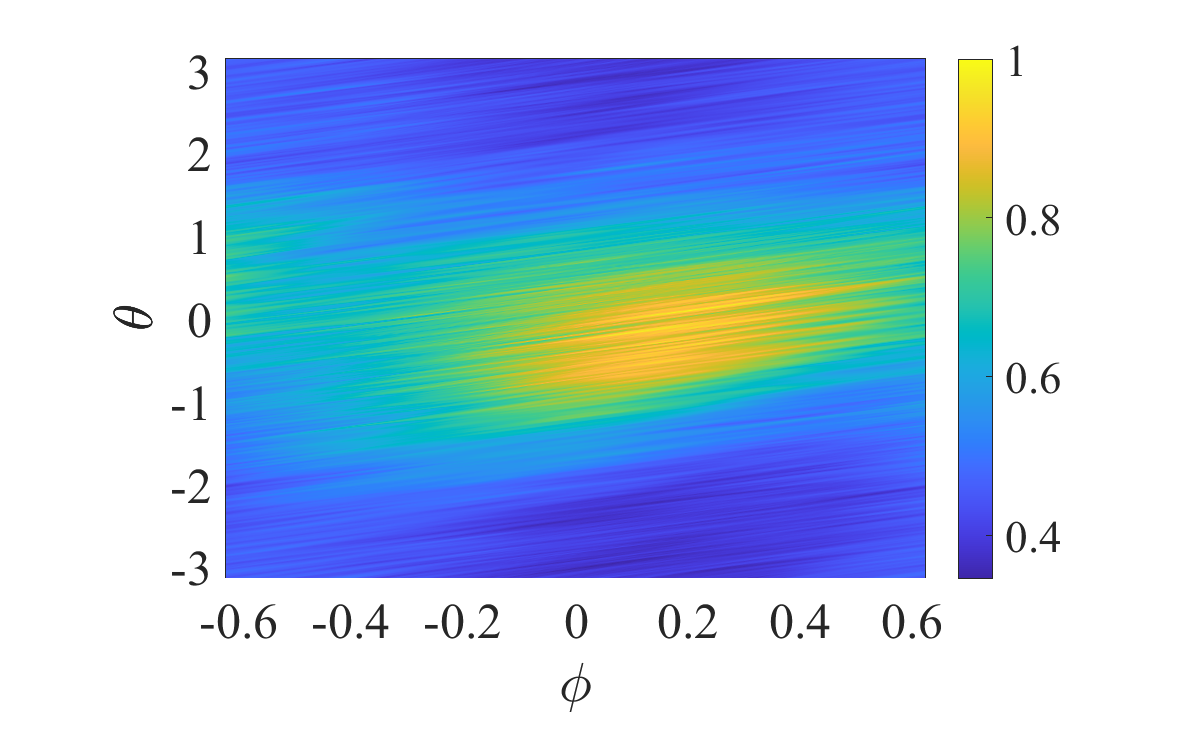}
 	\includegraphics[height=3.8cm,  trim=83 10 50 23, clip, angle=0]{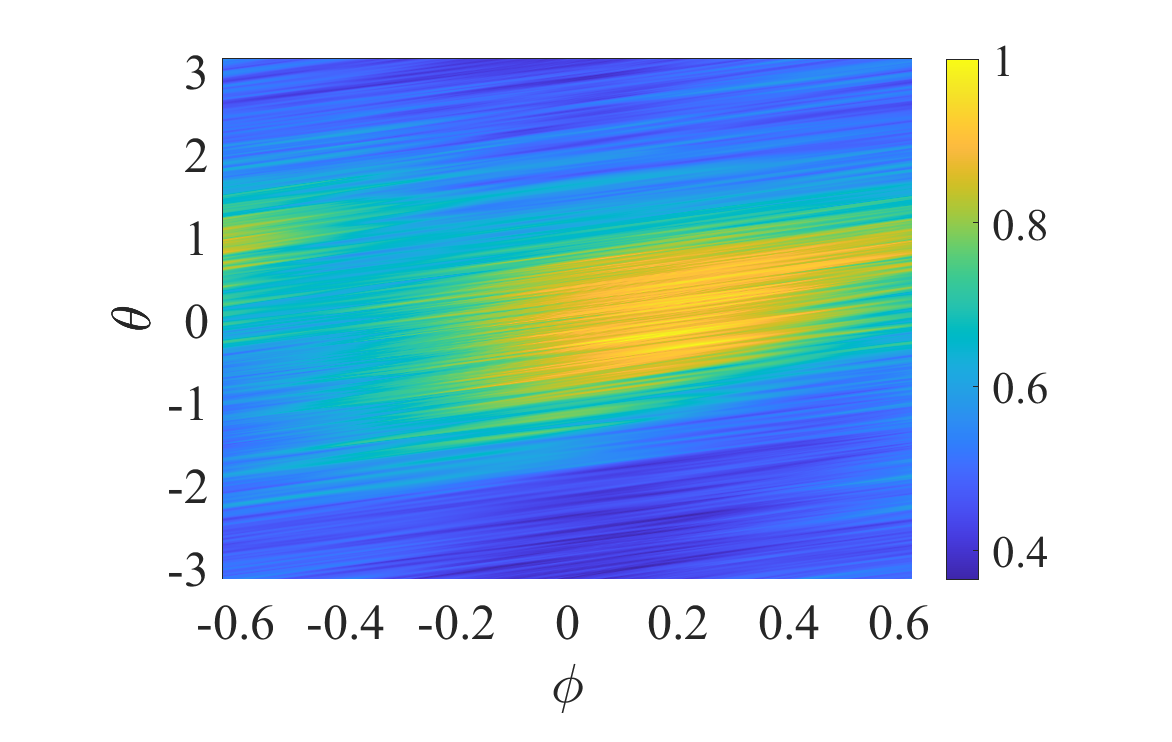}\\
 	\caption{Amplitude of the density (normalized to its maximum value) at the flux surface at $r/a=0.5$ vs toroidal and poloidal angles for a set of linear simulations in the W7-X configuration with EUTERPE (top) and GENE-3D (bottom) including several values of radial electric field, $Er=0, 20~\rm{kV/m},-20~\rm{kV/m}$, from left to right. }
 	\label{fig:LocalW7XwErNL}
 \end{figure}
 %*******************************************
 The displacement of the maximum instability with the electric field shown in Figure \ref{fig:LocalW7XwEr} is not observed in the nonlinear saturated phase. Only a very slight poloidal shift is hardly appreciated in the middle and right top panels corresponding to the EUTERPE simulations including an electric field. In the simulations with GENE-3D the shift is even harder to appreciate. This lack of displacement of maximum fluctuations region with the electric field is in contrast with that reported in full surface simulations \cite{Xanthopoulos20}, which show strong localization of density fluctuations and clear displacement of the regions of maximum amplitude of fluctuations when an electric field with Mach number similar to those considered here is included in the simulation. Note that the reference ion temperature in this work is larger than that in Ref. \cite{Xanthopoulos20}. The Mach number in our case with $|E_r|=20\rm{kV/m}$ is $M\sim0.01$.
 
 Now we will return to a more quantitative comparison and will study the influence of a radial electric field 
 on the turbulent heat transport.

 %*******************************************
 
 \subsection{Influence of Er on turbulent heat flux}\label{subSecElecFieldTurbHeatF}
 
 %*******************************************
 
 Finally, we devote this section to study the influence of a background radial electric field on the turbulent transport.
 For the nonlinear simulations with EUTERPE and GENE-3D including a radial electric field with strengths  $E_r=20~\rm{kV/m}$, $E_r=-20~\rm{kV/m}$ described in the previous sections and another one with and $E_r=40~\rm{kV/m}$, we computed the radial profile of the full-surface-integrated turbulent heat flux for both codes and compare them in Figure \ref{fig:HeatFluxW7XwEr}. The results for the simulations without electric field from Figure \ref{fig:heatFluxesW7X} are also shown here for comparison. A very slight reduction (within error bars) of the heat flux with respect to the reference case without $E_r$ is observed when the electric field is included. Again, this small reduction of the turbulent heat flux with $E_r$ contrasts with the results previously reported in \cite{Xanthopoulos20}, where a strong reduction of the ion heat diffusivity is observed when a radial electric field with Mach number $\rm{M}\sim 0.008$ (comparable to that in our simulations with $|E_r|=20~\rm{kV/m}$, with $\rm{M}\sim 0.01$) is included in the simulation.
 
 %*******************************************
 \begin{figure}[!h]
 	\centering
 	\includegraphics[width=7.5cm,  trim=30 10 50 0, clip, angle=0]{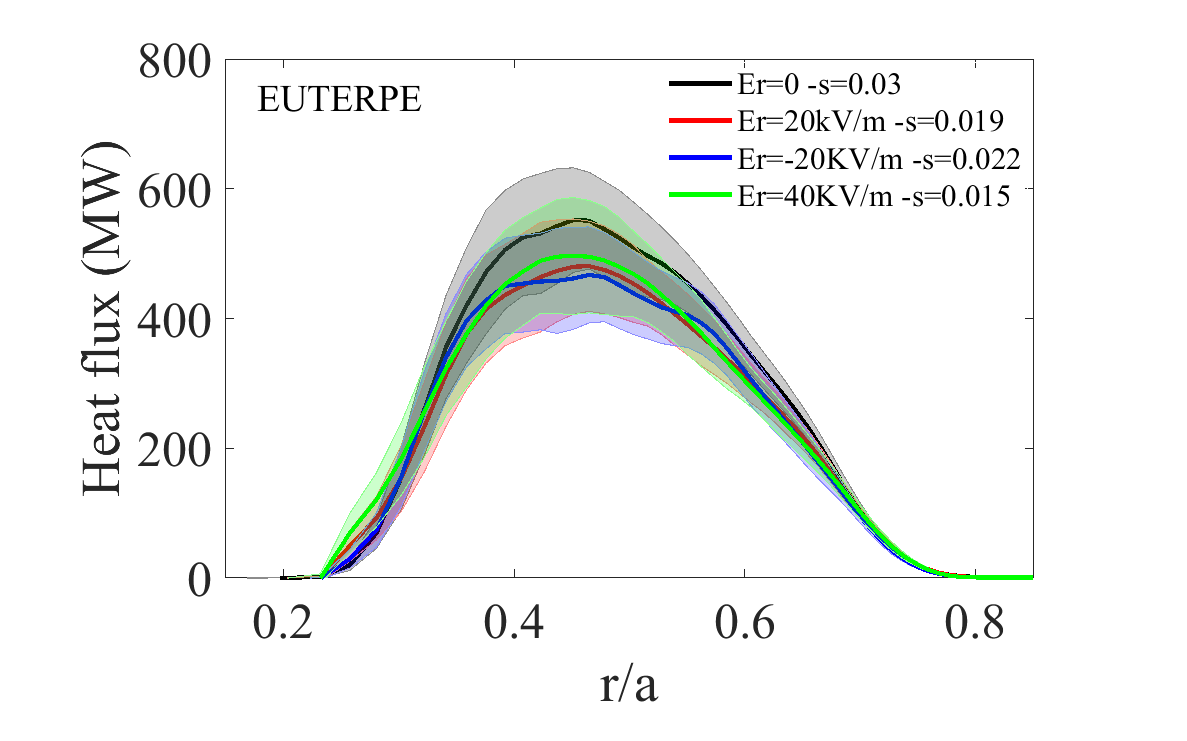}
 	\includegraphics[width=7.5cm,  trim=30 10 50 0, clip, angle=0]{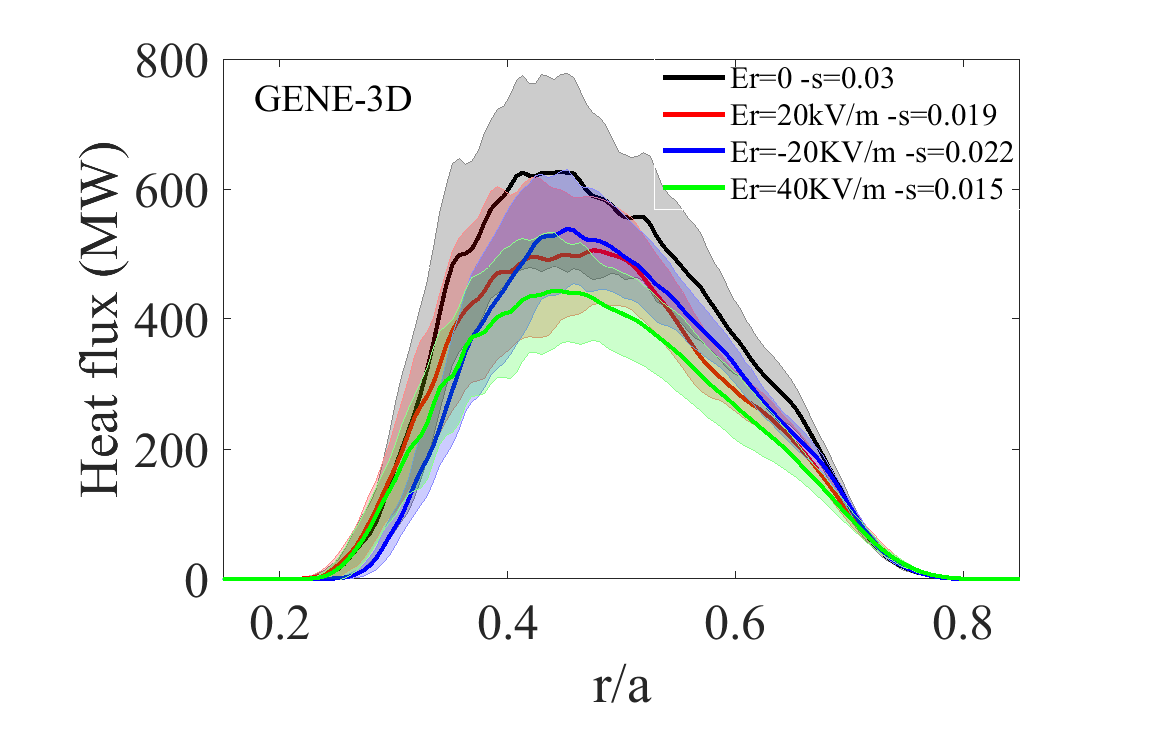}
 	\caption{Turbulent ion heat flux across the full surface versus the radial position for nonlinear simulations for the W7-X configuration with EUTERPE (left) and GENE-3D (right) including an electric field with different strengths $E_r=20~\rm{kV/m}$, $E_r=40~\rm{kV/m}$ and $E_r=-20~\rm{kV/m}$. The heat fluxes without electric field, shown in Figure \ref{fig:heatFluxesW7X} are also shown here for comparison.}
 	\label{fig:HeatFluxW7XwEr}
 \end{figure}
 %*******************************************
 {Contrary to the linear stabilization by $E_r$, the reduction of heat flux is almost independent on the sign of the electric field $E_r=20~\rm{kV/m}$ or $E_r=-20~\rm{kV/m}$. This is consistent with the fact that the linear instabilities are much more localized than the fluctuations in saturated turbulence, as discussed in sections \ref{secLocalizationInstab} and \ref{secLocalizationTurb}. 
 	The linear instabilities are strongly affected by local magnetic geometry, while global structures, such as the zonal flows, have an effect in the saturation of turbulence as well. A long-wavelength background electric field can displace the location of maximum instabilities \cite{Riemann16}, but its effect on the saturated turbulence fluctuations is diminished, as shown in Section \ref{subSecElecFieldTurb}, and the same can be expected for the turbulent transport derived from them.}
 
 In the EUTERPE case, increasing $E_r$ from $20~\rm{kV/m}$  to $40~\rm{kV/m}$ does not reduce the 
 heat flux further, in contrast with the reduction of growth rate in linear simulations. As for the case of GENE-3D, a slight reduction of heat flux is observed for the case with $E_r=40~\rm{kV/m}$ as compared to the case $E_r=20~\rm{kV/m}$. The reduction is small ($\sim12\%$), however.
 
 %*******************************************
 
 \section{Summary and conclusions}\label{secSumConcl}
 
 In this work, a thorough comparison of EUTERPE and GENE-3D codes has been carried out in magnetic configurations of LHD and W7-X. Three different settings have been chosen: linear simulations with adiabatic electrons, linear multi-species simulations with kinetic ions and electrons and nonlinear simulations with adiabatic electrons. 
 Besides the comparison of codes in these settings, two specific aspects that cannot be explored in local flux tube simulations have been studied in detail  in the simulations with adiabatic electrons, the localization of instabilities and turbulence over the flux surface and the influence of an electric field on localization, stabilization of linearly unstable modes and turbulent ion heat transport.
 
 In linear simulations with adiabatic electrons, excellent agreement between the codes was found  in LHD with respect to the mode number of the most unstable mode, its mode structure and spatial localization, both in the radial direction and over the flux surface. In W7-X the agreement in single simulations is not as good as in LHD but good agreement was also found in this configuration when the curves $\gamma -n$ are constructed from a set of simulations with increasing resolutions.   
 In linear simulations with kinetic electrons in W7-X, good agreement on the most unstable modes is found between the codes and a stabilization of small-scale modes is observed consistently by both codes in kinetic-electron simulations as compared to adiabatic-electron cases.
 
 Nonlinear simulation with adiabatic electrons in LHD using narrow profiles showed a strong dependence of the turbulent heat flux with the strength of the heating source used to sustain the kinetic profiles. Very good agreement between the codes on the turbulent heat flux was found for several values of the heating source strength in these simulations. In W7-X, the agreement between codes in the turbulent heat flux is also very good; both codes agree within error bars.
 
 With respect to the localization of instabilities in linear simulations, a clear localization of the region of maximum instability over the flux surface is found in both LHD and W7-X and consistently in both codes. The localization is significantly stronger in W7-X than in LHD. This result is in agreement with those previously reported in \cite{Nadeem01, Kornilov04, Sanchez19, Riemann16}. However, in nonlinear simulations in W7-X, the localization of turbulent fluctuations is largely diminished as compared to that in linear simulations, in contrast to previously reported results from full-surface simulations \cite{Xanthopoulos14}.
 
 A background radial electric field has been found to produce a  shift of the localization of maximum instability in linear simulations, in agreement with previous results \cite{Sanchez16, Riemann16}, however, its effect on the localization of maximum turbulent fluctuations is found to be almost negligible, again in contrast with previous full-surface results \cite{Xanthopoulos20}.  
 A radial electric field  produces a stabilization of linearly unstable modes, in agreement with previous results \cite{Riemann16}. This result contrasts with the situation in a tokamak, where a constant radial electric field does not stabilize the linear modes but only produces a Doppler shift. 
 A small reduction of turbulent transport by a background electric field, constant in radius, has been found by both codes in nonlinear simulations in W7-X including the electric field. Comparable reduction of heat flux is found in both codes, with small dependence on the sign of the electric field. A reduction of heat flux increasing with electric field strength is observed in GENE-3D, while in EUTERPE a weaker dependence with the field strength is observed. It remains for a future works studying the influence on the turbulent transport of more realistic sheared electric fields, such as those considered in \cite{EstradaNF2021,Fu2021}.

%%***************************************************************************************
\section{Acknowledgments}
%%***************************************************************************************

 Part of the simulations were carried out using the Marconi supercomputer at CINECA, from the EUROfusion infraestructure in the framework of the project TSVV-13. Some of the simulations were carried out using the Mare Nostrum-IV supercomputer. We acknowledge the computer resources, technical expertise and assistance provided by the Barcelona Supercomputing Center.
The work has been partially funded by the Ministerio de Ciencia, Innovaci\'on y Universidades of Spain  under project PGC2018-095307-B-I00.
This work has been carried out within the framework of the EUROfusion Consortium, funded by the European Union via the Euratom Research and Training Programme (Grant Agreement No 101052200-EUROfusion). Views and opinions expressed are however those of the authors only and do not necessarily reflect those of the European Union or the European Commission. Neither the European Union nor the European Commission can be held responsible for them.
%

%%***************************************************************************************
\section{References}
%%***************************************************************************************

\end{document}